\documentclass[a4paper,11pt]{article}
\pdfoutput=1
\usepackage{jheppub}
\usepackage[T1]{fontenc}
\usepackage[utf8]{inputenc}
\usepackage{amsmath,amssymb,bm,mathtools}
\usepackage{booktabs,tabularx,array,subcaption,microtype,orcidlink,placeins,needspace}
\graphicspath{{./}}
\newcommand{\Tr}{\operatorname{Tr}}
\newcommand{\Imn}{\operatorname{Im}}
\newcommand{\Ren}{\operatorname{Re}}
\newcommand{\diag}{\operatorname{diag}}
\newcommand{\id}{\mathbf 1}
\newcommand{\dd}{\mathrm d}
\newcommand{\GeV}{\mathrm{GeV}}
\newcommand{\eV}{\mathrm{eV}}
\newcommand{\Mpl}{M_{\rm Pl}}
\newcommand{\eqr}[1]{Eq.~\eqref{#1}}
\newcommand{\FS}{F_{\rm S}}
\newcommand{\GS}{G_{\rm S}}
\newcommand{\FV}{F_{\rm V}}
\newcommand{\eps}{\varepsilon}
\newcommand{\norm}[1]{\left\lVert#1\right\rVert_{\rm F}}
\numberwithin{equation}{section}
\hypersetup{pdftitle={Mixed leptogenesis and limits on neutrino-portal cancellation},pdfauthor={Yingnan Xu}}
\title{Mixed leptogenesis and limits on neutrino-portal cancellation}
\author[a,b]{Yingnan Xu\orcidlink{0009-0005-8296-0850}}
\affiliation[a]{Zhongtai Securities Institute for Financial Studies, Shandong University, Jinan, Shandong 250014, China}
\affiliation[b]{Department of Physics, Southern Methodist University, Dallas, Texas 75206, USA}
\emailAdd{yingnanx@mail.smu.edu}
\abstract{Heavy Majorana fermions can generate a lepton asymmetry and mediate a neutrino portal through the same visible and hidden couplings. We determine how a nonzero inclusive mixed lepton-number-violating source restricts cancellation of the matched portal. For two separated heavy mass levels, the imaginary overlap of the threshold amplitudes yields a lower bound on the portal norm in terms of the normalized asymmetry and heavy masses, together with a sharp domain at fixed threshold strengths. A flavour realization isolates this source while reproducing two nonzero light-neutrino masses and relates the portal to vacuum neutrino decay. On a specified reheating background, the same operator produces a scalar population whose inverse-energy moment inherits the portal bound. In the massless Born approximation, bath-produced scalars carry $4.81\times10^{-5}$ of the primary scalar energy but contribute $11.1$ times its variance at the benchmark. We evaluate the response to a prescribed scalar mass and derive conserving Hartree constraints. These results connect the mixed source to matching and production; cosmological viability requires subsequent coupled evolution, with the scalar potential and primordial state supplied independently.}
\begin{document}
\maketitle
\flushbottom

\widowpenalty=10000
\clubpenalty=10000
\displaywidowpenalty=10000

\section{Introduction}\label{sec:intro}
Heavy-particle contributions to a low-energy interaction can cancel even when the individual couplings are substantial. When the same amplitudes generate a CP-violating decay asymmetry, their relative phases also control the source. We study the resulting restriction in mixed visible--hidden leptogenesis: how small can the low-energy neutrino portal become while the heavy decays retain a nonzero inclusive mixed lepton-number-violating source? This question separates a connection enforced by heavy-particle matching from consequences that require additional flavour and cosmological assumptions.

Decay leptogenesis combines heavy Majorana neutrinos, the seesaw mechanism and electroweak sphaleron conversion~\cite{Sakharov:1967dj,Fukugita:1986hr,Minkowski:1977sc,Gell-Mann:1979vob,Ramond:1979family,Yanagida:1980xy,Mohapatra:1979ia,tHooft:1976rip,Manton:1983nd,Klinkhamer:1984di,Kuzmin:1985mm}. Its CP-violating amplitudes, neutrino-mass bounds and flavour evolution establish the relation between heavy decays and the surviving asymmetry~\cite{Weinberg:1979bt,Nanopoulos:1979gx,Luty:1992un,Flanz:1994yx,Covi:1996wh,Davidson:2002qv,Buchmuller:2004nz,Buchmuller:2005eh,Davidson:2008bu,Fong:2012buy,Blanchet:2012bk}. When the heavy particles also couple to a light fermion--scalar pair, the asymmetry receives a mixed self-energy contribution. The on-shell pair supplies the absorptive cut; off-shell heavy exchange generates a dimension-five interaction between the two light sectors. These two amplitudes contain the same Yukawa contractions.

The visible--hidden framework and its late interactions are established in earlier work. Falkowski, Ruderman and Volansky derived mixed decay asymmetries and matched a dimension-five portal giving a late scalar decay~\cite{Falkowski:2011}; related heavy-neutrino models also connect visible asymmetry production to a symmetric hidden population~\cite{Falkowski:2017lit,HerreroGarcia:2024seesaw}. Scalar-assisted sources with a vanishing conventional contribution occur in the construction of Barreiros et al.~\cite{Barreiros:2022fpi}. Off-shell heavy exchange drives collision leptogenesis in ref.~\cite{Bento:2001hidden}, and a mixed dimension-five interaction accompanies the Weinberg operator in the effective theory of Bla\v{z}ek et al.~\cite{Blazek:2025effective}. We derive the restriction that the inclusive mixed lepton-number-violating source places on cancellation of the matched coefficient, and follow its consequences for flavour-dependent widths and the scalar production variance.

The cancellation restriction follows from the two roles of the threshold amplitudes. With $T_I=Y_I^T\lambda_I$, the low-energy coefficient is their coherent sum, whereas the inclusive mixed source selects their imaginary overlap:
\[
 C=\frac{T_1}{M_1}+\frac{T_2}{M_2},
 \qquad J_+=\Imn\Tr(T_1T_2^\dagger)
          =M_2\Imn\Tr(T_1C^\dagger).
\]
Vanishing $C$ would require oppositely aligned threshold matrices with a real relative coefficient, which also sets $J_+=0$. Expressing decay asymmetries through effective coefficients is familiar from leptogenesis bounds and their flavour extensions~\cite{Davidson:2002qv,Antusch:2009nonunitary}. Here the rectangular visible--hidden coefficient gives, for two separated heavy mass levels,
\[
 \norm C\ge\frac{16\pi(1-r^2)}{M_1}|\eps_+|,
 \qquad r=M_1/M_2.
\]
The asymmetry $\eps_+$ is normalized per total decay of an equally populated incoherent lighter ensemble with an isotropic total-width block. The coefficient is generated by the displayed heavy exchange, at the matching and source orders specified in Section~\ref{sec:bound}. The allowed norm domain at fixed threshold strengths quantifies the remaining cancellation freedom. Independent matching contributions and additional heavy levels have the separate qualifications given there. Conventional visible, mixed lepton-number-conserving and flavour-traceless sources involve different contractions.

To connect the bounded source to the full inclusive asymmetry, we construct a flavour texture that removes the other inclusive contributions while retaining two nonzero light-neutrino masses. Flavour alignment and residual CP conditions can suppress conventional leptogenesis~\cite{Choubey:2010up,Chen:2016ptr,Li:2017zmk}. Under the separate visible-kernel cancellation conditions used here, one heavy field at each of two levels with a nonzero mixed source would force a rank-one light mass matrix. Two degenerate pairs permit an analytic realization with two massive light neutrinos. The construction then relates the mixed coefficient to their vacuum decay couplings. Its extension varies the source and the distribution of coupling strength at fixed light masses.

Reheating supplies the injection and entropy normalization needed to relate the mixed source to the baryon yield. We use perturbative inflaton decay with prompt heavy-particle decays, within the regimes distinguished from nonperturbative production in refs.~\cite{Dolgov:1982th,Abbott:1982hn,Traschen:1990sw,Kofman:1994rk,Shtanov:1994ce,Kofman:1997yn,GreeneKofman:2000,Allahverdi:2010xz,Lozanov:2019jxc}. The thermal history controls dilution and out-of-equilibrium production~\cite{Kolb:1990vq,Giudice:2000ex}, including the production and flavour evolution of the asymmetry~\cite{HahnWoernle:2008reheating,Datta:2022reheating,Datta:2023flavor,Zhang:2023nonthermal}. Eliminating the portal normalizations at fixed baryon yield and primary hidden radiation fraction yields a conditional lower envelope for the vacuum widths. The matched interaction also produces much softer scalars from the visible bath. Their inverse-energy moment can dominate scalar dispersion even when their energy density is small, linking the cancellation bound to a positive production functional~\cite{Du:2021distribution,Mukaida:2015density}.

The production spectra define a reference state for the subsequent response. Stimulated neutrino decay and thermal occupation effects are well established~\cite{Kaiser:1993lasing,Starkman:1993lasing,Kiessig:2010thermal,PerezGonzalez:2026thermal}. We specialize the finite-mass decay and inverse-decay framework to fixed Fermi--Dirac parents and derive the scalar moment constraints within a conserving Hartree closure. The scalar mass, quartic, Higgs portal and primordial state enter as independent inputs. Evolving the coupled distributions and their angular perturbations is a further step needed to infer cosmological free streaming~\cite{Escudero:2019decay,Barenboim:2020decay,Chen:2022lifetime}.

Section~\ref{sec:model} establishes the source--portal bounds. Section~\ref{sec:texture} gives the flavour realization and its decay consequences. Section~\ref{sec:reheat} imposes baryon normalization and derives the scalar production moments; Section~\ref{sec:numbers} evaluates these relations at a benchmark. Section~\ref{sec:medium} treats scalar response and the domain of the kinetic approximation. Appendix~\ref{app:cut} supplies the cut and spectrum derivations, production checks and scalar matching terms. The remaining appendices develop the flavour extensions and phase-selection example.

\section{The mixed source and the portal constraint}\label{sec:model}
\subsection{Fields, widths and heavy-exchange matching}
The heavy fields $X_{I,a}$ are left-handed Weyl fields with positive Majorana mass matrices $M_I\id_2$, where $I=1,2$, $a=1,2$ and $M_2>M_1$. Two massless Weyl fields $n_r$ and a complex scalar $S$ carry charges $+1$ and $-1$ under a conserved $U(1)_D$. The Standard Model and the heavy fields have zero hidden charge. In a canonical basis,
\begin{equation}
 {\cal L}_{\rm int}=-\sum_{I=1,2}
 \left[\frac12 M_I X_I^T X_I+X_I^T Y_I(L\cdot H)
       +\sqrt2 X_I^T\lambda_I n S\right]+\mathrm{h.c.}
 \label{eq:lag}
\end{equation}
Here $Y_I$ and $\lambda_I$ have dimensions $2\times3$ and $2\times2$.
The factor $\sqrt2$ fixes the hidden-coupling convention; the light theory contains one interacting complex scalar. The fermions and scalar obey Fermi--Dirac and Bose--Einstein statistics, respectively. Internal heavy multiplicity labels flavour. We use $v=174\,\GeV$ and the reduced Planck mass $\Mpl=2.435\times10^{18}\,\GeV$.

Define $K_{IJ}=Y_IY_J^\dagger$ and $D_{IJ}=\lambda_I\lambda_J^\dagger$. For massless decay products the widths of an individual heavy component, including charge-conjugate channels, are
\begin{equation}
 \Gamma^{\rm vis}_{I,a}=\frac{M_I}{8\pi}(K_{II})_{aa},
 \qquad
 \Gamma^{D}_{I,a}=\frac{M_I}{8\pi}(D_{II})_{aa}.
 \label{eq:widthheavy}
\end{equation}
The factor of two from the hidden coupling matches the visible weak-doublet multiplicity. For an equally populated incoherent pair the average total width is $\Gamma_I^{\rm tot}=M_I\Tr(K_{II}+D_{II})/(16\pi)$.

Integrating out the fields in \eqr{eq:lag} gives
\begin{equation}
 \begin{aligned}
 \Delta{\cal L}_5={}&\frac12 C_{LL}^{\alpha\beta}(L_\alpha\cdot H)(L_\beta\cdot H)
 +\sqrt2 C_{Ln}^{\alpha r}(L_\alpha\cdot H)n_rS
 +C_{nn}^{rs}n_rn_sS^2+\mathrm{h.c.},\\
 C_{LL}={}&\sum_I\frac{Y_I^TY_I}{M_I},\qquad
 C_{Ln}=\sum_I\frac{Y_I^T\lambda_I}{M_I},\qquad
 C_{nn}=\sum_I\frac{\lambda_I^T\lambda_I}{M_I}.
 \end{aligned}
 \label{eq:matching}
\end{equation}
These coefficients follow by substituting $X_I=-[Y_I(L\cdot H)+\sqrt2\lambda_InS]/M_I$ into the action. The visible term is the Weinberg operator~\cite{Weinberg:1979operator}; the other two coefficients follow from the same substitution for the present charged light fields. Every displayed term conserves $U(1)_D$.

\begin{figure}[htbp]
 \centering
 \begin{subfigure}{0.49\linewidth}
  \includegraphics[width=\linewidth]{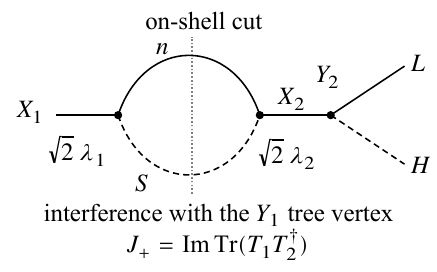}
  \caption{Mixed absorptive contribution.}\label{fig:mixedcut}
 \end{subfigure}\hfill
 \begin{subfigure}{0.49\linewidth}
  \includegraphics[width=\linewidth]{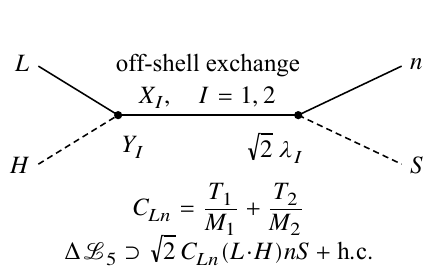}
  \caption{Tree-level portal matching.}\label{fig:portalmatching}
 \end{subfigure}
 \caption{Shared heavy amplitudes in the mixed source and low-energy matching,
with $T_I=Y_I^T\lambda_I$. (a) The on-shell $nS$ cut supplies an absorptive
self-energy correction that interferes with the visible tree vertex.
The displayed $J_+$ selects the inclusive mixed lepton-number-violating
contraction after summing the corresponding Majorana flows.
(b) Off-shell heavy exchange generates the mixed dimension-five operator.
Solid and dashed lines denote fermions and scalars, respectively.
Vertex labels denote the Lagrangian coupling matrices; flavour indices
and flow-dependent conjugations are implicit. The external labels name
fields, with crossed reactions and charge-conjugate diagrams understood.}
 \label{fig:mechanism}
\end{figure}

The heavy exchange has two roles, shown in Figure~\ref{fig:mechanism}.
An on-shell $nS$ intermediate state supplies the absorptive self-energy
that interferes with the visible decay amplitude. Off shell, the same
heavy exchange generates the local interaction $(L\cdot H)nS$.
The source probes the imaginary overlap of the threshold amplitudes,
whereas the matched coefficient sums them coherently. The relative phases
responsible for the source therefore restrict cancellation of the portal.
Section~\ref{sec:bound} quantifies this restriction.

The renormalized scalar mass and potential couplings are independent matching inputs. Section~\ref{sec:running} states their threshold sensitivity and the choices used in the decay and production calculations.

\subsection{The inclusive source and cancellation bound}\label{sec:bound}
Summing over visible flavours isolates the overlap relevant to portal cancellation. Define the threshold matrices $T_I=Y_I^T\lambda_I$ and write $C=C_{Ln}$. The relevant CP-odd invariant and its contribution to the asymmetry per total decay are
\begin{equation}
 \begin{aligned}
 J_+&=\Imn\Tr(K_{12}D_{12}^T)=\Imn\Tr(T_1T_2^\dagger),\\
 \eps_+&=\frac{\FS J_+}{8\pi\Tr\mathsf H_{11}},\qquad
 \mathsf H_{11}=K_{11}+D_{11}.
 \end{aligned}
 \label{eq:inclusivemixed}
\end{equation}
Here $\FS=\sqrt{x}/(1-x)$, with $x=M_2^2/M_1^2$, and $\eps_+$ denotes only the inclusive mixed lepton-number-violating contribution. The normalization assumes an equally populated incoherent lighter pair and an isotropic total-width block $\mathsf H_{11}$. We work with separated levels, $|M_2-M_1|\gg\Gamma_I^{\rm tot}$, so the nonresonant loop function applies~\cite{Covi:1996wh,Pilaftsis:1997jf,Pilaftsis:2003gt}. The physical $nS$ cut fixes the same coupling normalization as the diagonal width in Eq.~\eqref{eq:widthheavy}; Appendix~\ref{app:heavycut} gives the cut calculation. The full flavour source, including its other contractions, enters the source-isolation construction in Section~\ref{sec:texture}.

The invariant $J_+$ is unchanged by independent real orthogonal rotations of the positive-mass Majorana blocks and changes sign under CP~\cite{Kobayashi:1973fv,Jarlskog:1985ht,Branco:1999fs,Branco:2001pq,Branco:2005jr}. The portal adds $T_1/M_1$ and $T_2/M_2$ coherently, whereas the source selects their imaginary overlap. Complete cancellation would require these matrices to have a real negative relative coefficient, which also removes that imaginary overlap. To quantify incomplete cancellation, define the Frobenius norm by $\norm A^2=\Tr(AA^\dagger)$. The matching relation gives
\begin{equation}
 J_+=\Imn\Tr(T_1T_2^\dagger)
 =M_2\Imn\Tr(T_1C^\dagger)
 =M_1\Imn\Tr(CT_2^\dagger).
 \label{eq:identity}
\end{equation}
The terms proportional to $\Tr(T_IT_I^\dagger)$ are real. Thus $C=0$ forces $J_+=0$ for heavy exchange through these two mass levels. This relates the inclusive lepton-number-violating source to the mixed operator, while flavour-traceless sources and independently matched operators carry additional information.

For nonzero threshold norms, Cauchy--Schwarz applied to \eqr{eq:identity} yields
\begin{equation}
 \norm{C}\ge
 \max\left\{\frac{|J_+|}{M_2\norm{T_1}},
             \frac{|J_+|}{M_1\norm{T_2}}\right\}.
 \label{eq:normbound}
\end{equation}
To express the restriction in terms of the normalized source, we bound the lighter threshold strength using its visible and hidden partial widths. Since
\begin{equation}
 \norm{T_1}\le\sqrt{\Tr K_{11}\Tr D_{11}}
 \le\frac12\Tr\mathsf H_{11},
 \label{eq:thresholdnorm}
\end{equation}
Eq.~\eqref{eq:identity}, with $r=M_1/M_2$, gives
\begin{equation}
 \boxed{\norm{C}\ge
 \frac{16\pi(1-r^2)}{M_1}|\eps_+|.}
 \label{eq:normalizedbound}
\end{equation}
Equation~\eqref{eq:normalizedbound} ties a nonzero source to a minimum heavy-exchange portal strength without fixing the heavier threshold norm. It holds at the displayed tree-matching and one-loop source orders. For separately isotropic lighter blocks,
$K_{11}=\kappa_1\id_2$ and $D_{11}=\eta\kappa_1\id_2$,
with $\kappa_1,\eta>0$, the parameter $\eta$ is the hidden-to-visible width ratio of the lighter block. Then $\norm{T_1}^2=2\eta\kappa_1^2$ strengthens the result to
\begin{equation}
 \norm{C}\ge\frac{8\sqrt2\pi(1-r^2)}{M_1}
 \frac{1+\eta}{\sqrt\eta}|\eps_+|
 \ge\frac{16\sqrt2\pi(1-r^2)}{M_1}|\eps_+|.
 \label{eq:isotropicbound}
\end{equation}
Fixing light masses and imposing perturbativity can further restrict
attainability. The fixed-threshold domain below retains the information
about relative threshold strengths that is discarded in
Eq.~\eqref{eq:thresholdnorm}.

The allowed domain can be characterized more sharply when the two norms are fixed. Write $A_I=T_I/M_I$, $a_I=\norm{A_I}$, $j=J_+/(M_1M_2)$ and $c=\norm{C}$. Since $\Tr(A_1A_2^\dagger)=u+ij$, $c^2=a_1^2+a_2^2+2u$ and $u^2+j^2\le a_1^2a_2^2$, one obtains
\begin{equation}
 \left[c^2-a_1^2-a_2^2\right]^2+4j^2\le4a_1^2a_2^2.
 \label{eq:domain}
\end{equation}
A specified $j$ is feasible only for $|j|\le a_1a_2$. In this domain the minimum is $c^2_{\min}=a_1^2+a_2^2-2\sqrt{a_1^2a_2^2-j^2}$. It is attained within the threshold-matrix domain by proportional matrices with the corresponding relative phase. If only $s=a_1^2+a_2^2$ is fixed, minimizing over the relative magnitudes gives $c^2\ge s-\sqrt{s^2-4j^2}$, with $|j|\le s/2$. Constraints on the underlying Yukawa couplings and light masses can restrict which points in this matrix domain are physically attained.

\begin{figure}[htbp]
 \centering
 \includegraphics[width=0.72\linewidth]{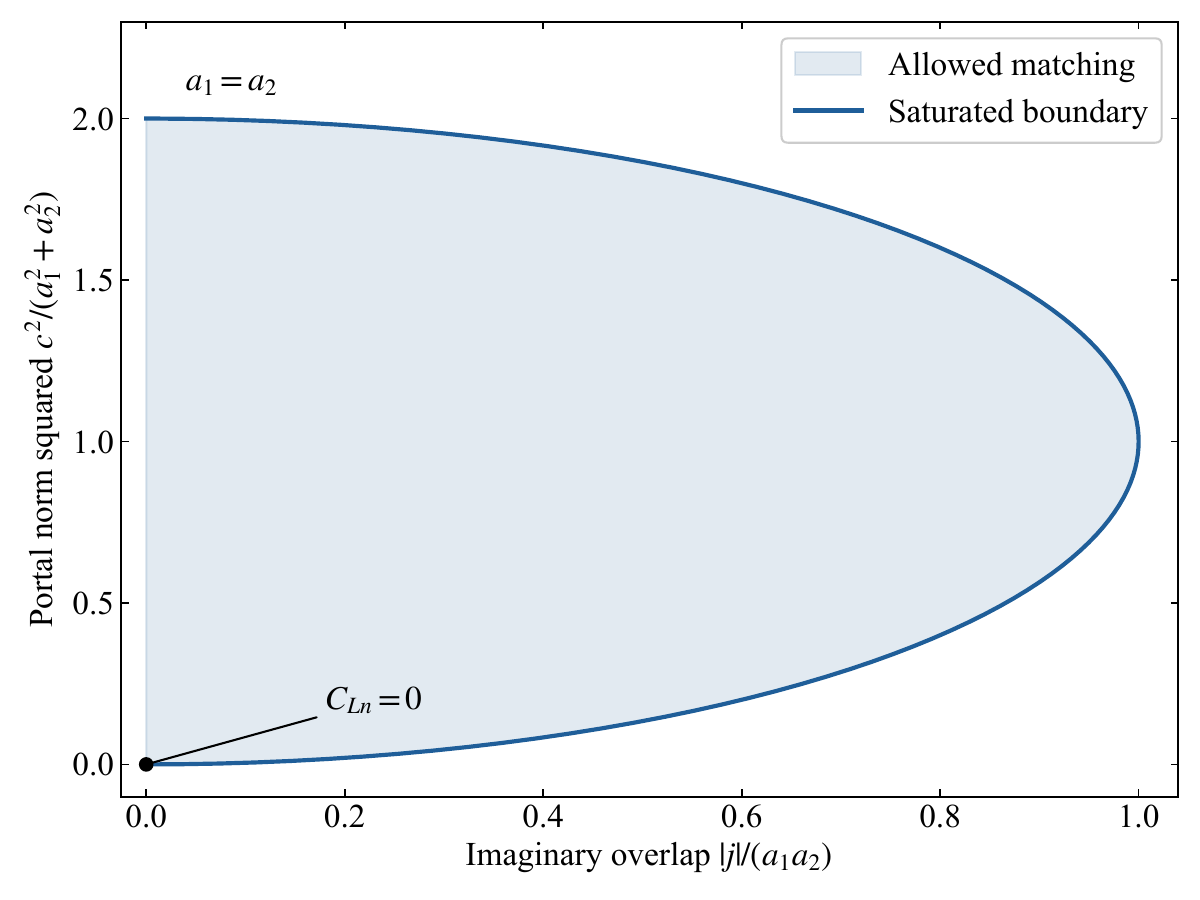}
 \caption{A mixed CP source restricts threshold cancellation. The allowed domain in Eq.~\eqref{eq:domain} is shown for $a_1=a_2$, with the portal norm squared normalized by $a_1^2+a_2^2$ and the imaginary overlap by $a_1a_2$. Converting the horizontal coordinate to the asymmetry per total decay also requires the masses and $\Tr\mathsf H_{11}$ in Eq.~\eqref{eq:inclusivemixed}. The boundary is a property of the threshold matrices before imposing a light-mass fit or perturbativity.}
 \label{fig:domain}
\end{figure}

Figure~\ref{fig:domain} shows how the imaginary overlap removes the strongly cancelled region at equal threshold strengths. Its source coordinate $|j|/(a_1a_2)$ is proportional to $|\eps_+|$ when the masses and total-width normalization are fixed. At maximal overlap the contributions are in quadrature, fixing $c^2=a_1^2+a_2^2$.

For an independent matching contribution $\Delta C$, the physical coefficient is $C_{\rm tot}=C+\Delta C$ and the bound becomes
\begin{equation}
 \norm{C_{\rm tot}}\ge
 \max\left\{0,\sqrt{a_1^2+a_2^2-2\sqrt{a_1^2a_2^2-j^2}}
                     -\norm{\Delta C}\right\}.
 \label{eq:thresholdbound}
\end{equation}
Cancellation therefore requires a threshold contribution at least as large as the heavy-exchange minimum. An unrestricted additional coefficient removes the positive lower bound on the physical portal. The same reverse-triangle subtraction applies to Eq.~\eqref{eq:normalizedbound}.

\subsection{Additional heavy thresholds}\label{sec:multilevel}
The two-level identity uses a single relative overlap. With additional heavier levels, distinct loop weights can retain information that the summed low-energy coefficient loses. The leading hierarchical relation still follows, with a finite-mass remainder that can be displayed directly. Let
$J_{1a}=\Imn\Tr(T_1T_a^\dagger)$, $r_a=M_1/M_a$ and
$C=\sum_{a\ge1}T_a/M_a$. Then
\begin{equation}
 \sum_{a>1}\FS(r_a)J_{1a}
 =-M_1\Imn\Tr(T_1C^\dagger)-M_1\sum_{a>1}\frac{r_a^2}{1-r_a^2}\frac{J_{1a}}{M_a}.
 \label{eq:multilevelremainder}
\end{equation}
This follows by separating
$\FS(r_a)=-r_a-r_a^3/(1-r_a^2)$.
At fixed amplitudes and source normalization, cancellation of $C$ removes
the leading hierarchical term; several heavier levels can leave a cubic
mass-ratio remainder. With one heavier level, Eq.~\eqref{eq:identity}
also removes that remainder. The dimension-five coefficient thus fixes the leading hierarchical source; the remainder retains information about the individual heavier levels.

\section{A vanishing visible source and light-neutrino decay}\label{sec:texture}
The general bound concerns one specified inclusive source. We now construct a flavour texture that makes this the full inclusive asymmetry while retaining two nonzero light-neutrino masses. The alignment removes the conventional visible source and the inclusive mixed lepton-number-conserving contribution. Matching the surviving portal then connects the bound to physical neutrino decay widths.

\subsection{Flavour source and the alignment condition}\label{sec:flavoursource}
To isolate the bounded contribution, we need the source as a flavour matrix as well as its trace. Within each positive degenerate Majorana block the residual basis transformations are real orthogonal. Under independent $O_I$, a lepton rotation $V_\ell$ and a hidden rotation $V_D$, the couplings transform as $Y_I\to O_IY_IV_\ell^\dagger$ and $\lambda_I\to O_I\lambda_IV_D^\dagger$. We use lepton-density matrices transforming as $\rho_\ell\to V_\ell\rho_\ell V_\ell^\dagger$. For $Q=K,D$, define
\begin{equation}
 \begin{aligned}
 A[Q]&=Y_1^TQ_{12}Y_2^*,&
 {\cal E}[Q]&=\left[\frac{A[Q]-A[Q]^\dagger}{2i}\right]^T,\\
 B[Q]&=Y_1^TQ_{12}^*Y_2^*,&
 {\cal P}[Q]&=\left[\frac{B[Q]-B[Q]^\dagger}{2i}\right]^T.
 \end{aligned}
 \label{eq:kernels}
\end{equation}
The transpose makes both Hermitian kernels transform in the representation of $Y_1^\dagger Y_1$. For an equally populated incoherent lighter pair with an isotropic total-width block $K_{11}+D_{11}$, the source per total decay is
\begin{equation}
 \eps=\frac{(\FV+\FS){\cal E}[K]+\GS{\cal P}[K]
                  +\FS{\cal E}[D]+\GS{\cal P}[D]}
                  {8\pi\Tr(K_{11}+D_{11})}.
 \label{eq:source}
\end{equation}
The additional loop functions are $\GS=1/(1-x)$ and $\FV=\sqrt{x}[1-(1+x)\ln(1+x^{-1})]$, with $x$ and $\FS$ defined in Section~\ref{sec:bound}. Taking the trace of the $\FS\mathcal E[D]$ term recovers $\eps_+$ in Eq.~\eqref{eq:inclusivemixed}.
The two Majorana-flow orientations of the $nS$ self-energy cut give $D_{12}$ and $D_{12}^*$. In the scalar-assisted construction of Barreiros et al.~\cite{Barreiros:2022fpi}, scalar couplings between heavy neutrinos supply additional decay sources. The present cut contains light charged $nS$ states; exchange of the same heavy levels generates the mixed operator among the surviving light fields.

The other inclusive mixed contraction is $J_-=\Imn\Tr(K_{12}D_{12}^\dagger)$, associated with $\mathcal P[D]$. Removing this contribution while retaining $J_+$ makes the cancellation bound applicable to the full inclusive mixed source. The real-orthogonal alignment $D_{12}^T=c_U O K_{12}O$, with $c_U>0$, gives $J_-=0$: setting $A=K_{12}O$ reduces the relevant trace to $c_U\Tr(AA^*)$, which is real. The texture below uses $O=\id_2$. This relation is imposed as a matching condition. It eliminates the inclusive mixed lepton-number-conserving contribution; its flavour-traceless kernel can remain nonzero.

\subsection{The aligned texture}
The light-mass requirement prevents the simplest rank-one realization of this source isolation. For one heavy field at each level, let $y_1,y_2$ be the visible rows and $z=y_1y_2^\dagger$. If $z=0$, the mixed invariant vanishes. If $z\ne0$, separate vanishing of the conventional lepton-number-violating kernel requires the rank-one matrix $z y_1^T y_2^*$ to be Hermitian. Its column space must then equal that of its adjoint, so $y_1$ and $y_2$ are proportional and $\operatorname{rank}C_{LL}\le1$. Thus one field at each level is insufficient under these source-isolation conditions. We use degenerate pairs with isotropic widths to accommodate the two massive light directions; this argument does not establish a universal minimum for other source textures. For a complex matrix $W$ with orthonormal rows, $WW^\dagger=\id_2$, take
\begin{equation}
 \begin{gathered}
 Y_1=\sqrt{\kappa_1}W,\qquad Y_2=\sqrt{\kappa_2}F^\dagger W,\qquad
 \lambda_1=\sqrt{\eta\kappa_1}\id_2,\qquad
 \lambda_2=c_U\sqrt{\frac{\kappa_2}{\eta}}F^*,\\
 F=\frac12\begin{pmatrix}1+i&1-i\\-1+i&-1-i\end{pmatrix},\\
 F^\dagger F=\id_2,\qquad F^2=i\id_2,\qquad
 FF^T=-\sigma_1,\qquad F^TF=\sigma_1 .
 \end{gathered}
 \label{eq:texture}
\end{equation}
Here $\kappa_I,\eta,c_U$ are positive. The orthonormal rows of $W$ fix a rank-two visible subspace, the unitary matrix $F$ preserves the isotropic widths, and the relation between $Y_2$ and $\lambda_2$ imposes the mixed-source alignment. In particular, $\eta=\Gamma^D_{1,a}/\Gamma^{\rm vis}_{1,a}$. The bilinears are
$K_{12}=\sqrt{\kappa_1\kappa_2}F$, $D_{12}=c_U\sqrt{\kappa_1\kappa_2}F^T$,
$K_{II}=\kappa_I\id_2$, $D_{11}=\eta\kappa_1\id_2$ and
$D_{22}=c_U^2\kappa_2\id_2/\eta$. Appendix~\ref{app:phase} supplies a discrete phase realization.

To fit the light masses and mixing, let $U$ denote the lepton mixing matrix and choose, for normal ordering,
\begin{equation}
 \begin{aligned}
 W&=O_HE_{23}U^\dagger,\\
 O_H&=\frac1{\sqrt2}\begin{pmatrix}1&1\\1&-1\end{pmatrix},\\
 E_{23}&=\begin{pmatrix}0&1&0\\0&0&1\end{pmatrix},\\
 \frac{\kappa_{1,2}}{M_{1,2}}&=\frac{m_3\pm m_2}{2v^2}.
 \end{aligned}
 \label{eq:seesaw}
\end{equation}
The upper sign belongs to level 1. Then $m_\nu=-v^2C_{LL}=-U^*\diag(0,m_2,m_3)U^\dagger$; a common Takagi phase makes the massive eigenvalues positive. The lightest neutrino is massless at this matching order. This choice specifies a subspace of the Casas--Ibarra parameterization~\cite{Casas:2001sr}.

Substituting \eqr{eq:texture} into \eqr{eq:kernels} gives
$A[K]=-\kappa_1\kappa_2W^T\sigma_1W^*$ and
$B[K]=\kappa_1\kappa_2W^TW^*$. Both are Hermitian, so
$\mathcal E[K]=\mathcal P[K]=0$ as matrices. The mixed contractions are
$A[D]=ic_U\kappa_1\kappa_2W^TW^*$ and
$B[D]=ic_U\kappa_1\kappa_2W^T\sigma_1W^*$.
With $r=M_1/M_2$, the resulting source is
\begin{equation}
 \eps=\frac{c_U\kappa_2\FS}{16\pi(1+\eta)}
       U\diag(0,1+r,1-r)U^\dagger,\qquad
 \eps_{\rm inc}\equiv\Tr\eps=
 -\frac{c_U}{1+\eta}\frac{M_1(m_3-m_2)}{16\pi v^2(1-r^2)} .
 \label{eq:epsclosed}
\end{equation}
The null eigenvector coincides with the massless-neutrino direction. The traceless self-energy contribution splits the two nonzero source eigenvalues by the heavy-mass ratio. Phases in $U$ enter the flavour components, while the inclusive source depends on the mass difference $m_3-m_2$. The mixed invariant is $J_+=2c_U\kappa_1\kappa_2$. Since $D_{12}=ic_UK_{12}^*$, this texture saturates $|J_+|\le\norm{K_{12}}\norm{D_{12}}$. This property concerns the relative visible and hidden couplings and complements the conventional visible-source mass bounds~\cite{Davidson:2002qv,Garbrecht:2024ngz}.

\subsection{Light matching and vacuum widths}
The source-isolating texture now fixes the low-energy interaction. Its mixed coefficient determines the decay amplitudes, while the visible coefficient retains the neutrino mass fit. The three coefficients in \eqr{eq:matching} reduce to
\begin{equation}
 \begin{aligned}
 C_{LL}&=W^T\left(\frac{\kappa_1}{M_1}\id_2-
                         \frac{\kappa_2}{M_2}\sigma_1\right)W,\\
 C_{Ln}&=bW^T,\qquad b=\frac{\sqrt\eta\,\kappa_1}{M_1}
                         -i\frac{c_U\kappa_2}{\sqrt\eta M_2},\\
 C_{nn}&=\frac{\eta\kappa_1}{M_1}\id_2+
                         \frac{c_U^2\kappa_2}{\eta M_2}\sigma_1 .
 \end{aligned}
 \label{eq:matchtexture}
\end{equation}
The different signs in $C_{LL}$ and $C_{nn}$ follow from $FF^T=-\sigma_1$ and $F^TF=\sigma_1$. In $C_{Ln}$ the two threshold contributions are in quadrature; this phase relation prevents a cancellation within the texture.

After electroweak symmetry breaking the coupling matrix is
$g_{ir}=\sqrt2 vb(U^TW^T)_{ir}$, up to the common Takagi phase.
Since $U^TW^T=E_{23}^TO_H^T$, the massless state has zero coupling and
$\sum_r|g_{2r}|^2=\sum_r|g_{3r}|^2\equiv g_\nu^2=2v^2|b|^2$.
A hidden-flavour rotation writes the interaction as $g_\nu\nu_i n_iS$ for $i=2,3$.
For massless decay products the inclusive Majorana width is
\begin{equation}
 \Gamma_i^{(0)}=\frac{m_i g_\nu^2}{16\pi}=\frac{m_i}{32\pi v^2}\left[\eta(m_3+m_2)^2+\frac{c_U^2}{\eta}(m_3-m_2)^2\right],
 \qquad i=2,3.
 \label{eq:lightwidth}
\end{equation}
The heavy masses cancel along the family with fixed light masses because \eqr{eq:seesaw} fixes the ratios of the Yukawa norms to the heavy scales. At fixed Yukawa couplings the heavy fields decouple. The widths for massless decay products obey $\Gamma_3^{(0)}/\Gamma_2^{(0)}=m_3/m_2$.

The relation between the operator norm and decay extends beyond equal row
norms. Let $G=\sqrt2vU^TC$, with the common Takagi phase understood.
When its nonzero rows lie in the massive-neutrino subspace and both daughter
particles are massless,
\begin{equation}
 \mathcal W_\nu\equiv\sum_{i\in\mathrm{massive}}
 \frac{\Gamma_i^{(0)}}{m_i}
 =\frac{v^2}{8\pi}\norm{C}^2
 \ge32\pi v^2\frac{(1-r^2)^2}{M_1^2}|\eps_+|^2.
 \label{eq:weightedwidth}
\end{equation}
A coupling to an exactly massless active mode contributes to the operator
norm but supplies no rest-frame decay width; the aligned rank-two texture
satisfies the stated support condition. Finite daughter masses require the
corresponding kinematic weights and open-channel restrictions. Individual
widths retain the flavour information in $GG^\dagger$.

The lifetime limits depend on which high-energy information is held fixed. With the light masses and $c_U$ fixed, define $m_+=m_3+m_2$ and $m_-=m_3-m_2$ and vary the hidden-to-visible width ratio $\eta>0$. Equation~\eqref{eq:lightwidth} gives
\begin{equation}
 \Gamma_i^{(0)}\ge\frac{m_ic_U(m_3^2-m_2^2)}{16\pi v^2},
 \qquad \eta_{\min}=\frac{c_U m_-}{m_+}.
 \label{eq:fixedcu}
\end{equation}
Reducing the lighter-level hidden coupling raises the heavier-level coupling required to retain the alignment. Their positive contributions balance at the minimum.

For fixed $M_1$, $r$ and inclusive source, \eqr{eq:epsclosed} instead fixes
$\alpha=c_U/(1+\eta)$. Substitution into \eqr{eq:lightwidth} yields
$(m_+^2+\alpha^2m_-^2)\eta+2\alpha^2m_-^2+\alpha^2m_-^2/\eta$ inside the brackets. Hence
\begin{equation}
 \Gamma_i^{(0)}\ge\frac{m_i\alpha m_-}{16\pi v^2}
 \left[\sqrt{m_+^2+\alpha^2m_-^2}+\alpha m_-\right],
 \qquad
 \eta_{\min}^{(\alpha)}=\frac{\alpha m_-}{\sqrt{m_+^2+\alpha^2m_-^2}}.
 \label{eq:fixedalpha}
\end{equation}
Equations~\eqref{eq:fixedcu} and \eqref{eq:fixedalpha} therefore minimize along different parameter trajectories. These algebraic minima precede any further perturbative or dynamical restrictions. A finite scalar mass introduces a separate limitation: even a nonzero portal gives no neutrino decay through a closed channel. For a common vacuum scalar mass $m_{S,0}$ and a massless hidden fermion,
\begin{equation}
 \Gamma_i(m_{S,0})=\Gamma_i^{(0)}
       \left(1-\frac{m_{S,0}^2}{m_i^2}\right)^2\Theta(m_i-m_{S,0}).
 \label{eq:scalarwidth}
\end{equation}
Both neutrinos decay for $m_{S,0}<m_2$, only $\nu_3$ has this two-body decay for $m_2<m_{S,0}<m_3$, and the crossed scalar decays open above the corresponding thresholds. The width bounds apply when both decay products are massless.

\subsection{The family with a vanishing visible source}
\label{sec:classification}

The aligned example has equal-strength orthogonal decay rows. To identify which features follow from the source isolation and which belong to this endpoint, we vary the internal matrix while preserving the light mass matrix. Within the aligned two-block portal ansatz of Eq.~\eqref{eq:texture}, allow $F$ to be any unitary matrix of dimension two. The orthonormality of the rows gives
$\mathcal P[K]=0$, while $\mathcal E[K]=0$ is equivalent to Hermiticity of
$R_F=FF^T$. This matrix is symmetric and unitary, so Hermiticity makes
it real and orthogonal. The choices $R_F=\pm\id_2$ give
$\Imn\Tr F^2=0$. A nonzero mixed inclusive source therefore requires the
reflection class, with eigenvalues $+1$ and $-1$.

A real orthogonal change of heavy basis brings this class to
$R_F=\sigma_3$. Every allowed $F$ then has the form
$F=\operatorname{diag}(1,-i)O$, where $O$ is real orthogonal. A representative
on the selected CP branch, with $0\le\theta\le\pi/2$, is
\begin{equation}
 F_\theta=\begin{pmatrix}1&0\\0&-i\end{pmatrix}
 \begin{pmatrix}\cos\theta&\sin\theta\\-\sin\theta&\cos\theta\end{pmatrix},
 \qquad \xi=\sin^2\theta,
 \qquad J_+=2c_U\kappa_1\kappa_2\xi .
 \label{eq:reflectionfamily}
\end{equation}
The other orthogonal component supplies the CP-conjugate source sign.
Writing $h_1=\kappa_1/M_1$ and $h_2=\kappa_2/M_2$, we obtain
$C_{LL}=W^T(h_1\id_2+h_2\sigma_3)W$. Thus varying $\xi$ preserves the
light-neutrino masses $m_3=v^2(h_1+h_2)$ and $m_2=v^2(h_1-h_2)$ for $h_1>h_2>0$.
The mixing matrix is retained by choosing the same light mass eigenvectors
for the rows of $W$. The inclusive asymmetry changes according to
$\eps_{\rm inc}(\xi)=\xi\eps_{\rm inc}(1)$.

The mixed coefficient becomes
$C_{Ln}=W^T[x\id_2+yF_\theta^{*2}]$, with
$x=\sqrt\eta\,h_1$ and $y=c_Uh_2/\sqrt\eta$. The squared coupling norms and vacuum widths for massless daughters in the light-neutrino mass basis are
\begin{equation}
 \begin{aligned}
 g_{3,2}^2&=2v^2\bigl[x^2+y^2\pm2xy(1-\xi)\bigr],\\
 \Gamma_{3,2}^{(0)}&=\frac{m_{3,2}}{32\pi v^2}
 \left[\eta m_+^2+\frac{c_U^2m_-^2}{\eta}
 \pm2c_U m_+ m_-(1-\xi)\right],
 \end{aligned}
 \label{eq:reflectionwidths}
\end{equation}
where the upper sign refers to $\nu_3$.
The source therefore measures the same continuous parameter that separates
the two inclusive coupling norms. At $\xi=1$, the representative obeys
$F_\theta^2=i\id_2$ and is related by a real orthogonal transformation to
the texture in Eq.~\eqref{eq:texture}. This endpoint maximizes the mixed
source for fixed portal norms and gives Eq.~\eqref{eq:lightwidth}.
At fixed $\eta$ and $c_U$, the sum $g_2^2+g_3^2$ is independent of $\xi$. The reflection trajectory lies at the centre $\norm C^2=a_1^2+a_2^2$ of Eq.~\eqref{eq:domain} and reaches its boundary at $\xi=1$. The full flavour source and singular couplings are given in Appendix~\ref{app:reflection}.
For the balanced choice $\eta=c_U m_-/m_+$, the ratios to the common endpoint
value are $g_3^2/g_0^2=2-\xi$ and $g_2^2/g_0^2=\xi$.

The coupling matrix also determines whether the two decays populate orthogonal hidden states. Let $G_{\rm m}$ be the $2\times2$ submatrix of $G$ formed by the massive rows in the order $(\nu_3,\nu_2)$. Then
\begin{equation}
 \begin{aligned}
 G_{\rm m}G_{\rm m}^\dagger&=2v^2\left[(x^2+y^2)\id_2+xy\mathcal H_\theta\right],\\
 \mathcal H_\theta&=
 \begin{pmatrix}
 2(1-\xi)&(1+i)\sin2\theta\\
 (1-i)\sin2\theta&-2(1-\xi)
 \end{pmatrix}.
 \end{aligned}
 \label{eq:reflectioncouplings}
\end{equation}
At $\xi=1$ the mixed source is maximal and the two massive-neutrino rows
have equal norms and zero overlap. For $0<\xi<1$, the rows overlap. A hidden basis change leaves $G_{\rm m}G_{\rm m}^\dagger$ unchanged and therefore preserves this overlap. At $\xi=0$ the overlap also vanishes, together with the inclusive mixed source. The independent-channel collision description used later is restricted to the maximal-source endpoint $\xi=1$.

\begin{figure}[htbp]
 \centering
 \includegraphics[width=0.72\linewidth]{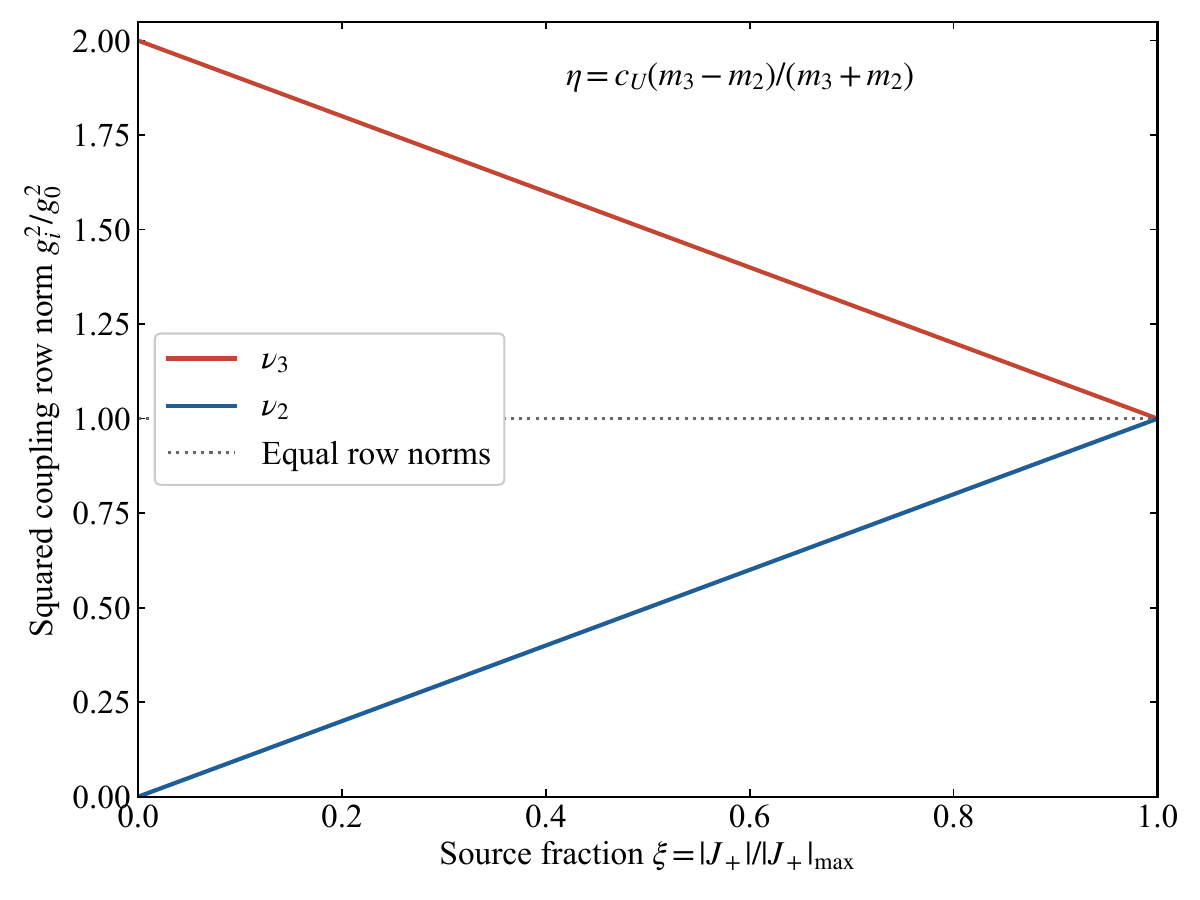}
 \caption{The reflection family redistributes the light-neutrino coupling at fixed masses and total portal norm. The squared row norms are normalized to their common endpoint value $g_0^2$, with $\eta=c_U(m_3-m_2)/(m_3+m_2)$. The conventional source vanishes throughout the family, and $\xi=\eps_{\rm inc}(\xi)/\eps_{\rm inc}(1)$. Equal-strength orthogonal rows occur at $\xi=1$; intermediate points also have the hidden-flavour overlap in Eq.~\eqref{eq:reflectioncouplings}.}
 \label{fig:reflection}
\end{figure}

Figure~\ref{fig:reflection} illustrates this redistribution at balanced threshold strengths. Decreasing $\xi$ lowers the inclusive source and transfers coupling strength from $\nu_2$ to $\nu_3$, while keeping the sum of their squared couplings fixed.

\FloatBarrier
\subsection{Scalar matching inputs and radiative scope}\label{sec:running}
The flavour construction fixes the decay coupling; the scalar matching parameters determine whether the light channels remain open. Before electroweak symmetry breaking, the scalar potential is $V_S=m_{S,\rm sym}^2S^\dagger S+\lambda_S(S^\dagger S)^2+\lambda_{HS}S^\dagger S H^\dagger H$, with $\langle S\rangle=0$. Its renormalized coefficients are independent matching parameters. In the tree-level matching used here, the vacuum scalar mass satisfies $m_{S,0}^2=m_{S,\rm sym}^2+\lambda_{HS}v^2$. The benchmark sets $m_{S,0}=0$ and $\lambda_{HS}=0$; the scalar self-coupling remains free in the finite-density analysis. The hidden charge forbids a Weyl Majorana mass and permits $S^\dagger S$. Consequently, an ultralight scalar requires cancellation of heavy-threshold contributions. Their characteristic scale is $\Tr(D_{22})M_2^2/(8\pi^2)$, of order $10^{24}\,\GeV^2$ at the benchmark. An open decay channel below $m_3\simeq0.05\,\eV$ corresponds to a mass-squared matching sensitivity of order $10^{-45}$. This estimate specifies the scalar mass choice; the flavour and width relations throughout use renormalized parameters at the heavy thresholds.

The same heavy sector generates scalar quartic and Higgs-portal thresholds. In the isolated heavier-block $\overline{\mathrm{MS}}$ effective potential at $\mu_R=M_2$, their contributions are $\Delta\lambda_S=-2.57\times10^{-2}$ and $\Delta\lambda_{HS}=-3.10\times10^{-4}$, derived in Appendix~\ref{app:quartic}. A negligible physical Higgs portal requires an independent cancellation, in addition to the scalar-mass choice. It otherwise contributes both an electroweak mass shift and Higgs-mediated scalar production. We impose $\lambda_{HS}=0$ in the leading-order effective theory used for production and the vacuum widths. The tiny positive $\lambda_S$ values in Section~\ref{sec:medium} are illustrative renormalized parameters probing the density response. The hidden charge and the discrete phase construction permit these scalar operators and provide no protection of their hierarchy. A pseudo-Nambu--Goldstone interpretation would require a symmetry-consistent completion and a new matching of its interactions.

The reflection family varies the mixed source while keeping the conventional source zero. A separate orientation deformation restores a visible contribution while preserving isotropic diagonal widths. Appendix~\ref{app:orientation} gives its source and the domain compatible with the same two light masses; the parameter $\delta=0$ recovers the source-isolating texture. This distinction matters when assessing possible resonant corrections.

The texture specifies canonical parameters in a common leading-order matching convention: the two tree contributions are summed just below $M_1$, and the coefficients are used without running in the low-energy widths. The measured light masses set this tree-level normalization. Sequential threshold matching and evolution of the Weinberg operator are discussed in ref.~\cite{Antusch:2005gp}. The mixed coefficient requires its own evolution, beyond the present tree-level calculation. The displayed numerical digits identify the benchmark parameters; loop corrections remain to be calculated.

Both unitary families preserve $K_{II}=\kappa_I\id_2$ and $D_{II}=d_I\id_2$, with $d_1=\eta\kappa_1$ and $d_2=c_U^2\kappa_2/\eta$. Writing $\mathsf H_{IJ}=K_{IJ}+D_{IJ}$, the mass equation gives
\begin{equation}
 \begin{aligned}
 16\pi^2\frac{\dd M}{\dd\ln\mu_R}&=\mathsf HM+M\mathsf H^T,\\
 16\pi^2\beta_{M,II}&=2M_I(\kappa_I+d_I)\id_2,\\
 16\pi^2\beta_{M,12}&=M_2\mathsf H_{12}+M_1\mathsf H_{12}^* .
 \end{aligned}
 \label{eq:massrunning}
\end{equation}
The initial diagonal contribution is a common shift within each pair, including for $\delta\ne0$. For separated mass levels, off-block mixing affects their mass eigenvalues at second order in that perturbation. Further Yukawa running or additional flavour spurions can generate anisotropy and splitting. When such perturbations produce interfering modes with splittings comparable to their widths and nonzero CP correlations, a resummed or density-matrix calculation is required~\cite{Pilaftsis:2003gt,BhupalDev:2014pfm}. The nonresonant formulas describe the stated isotropic-block family and the separated levels.

At the benchmark, the heavier-block loop factor is $(\kappa_2+d_2)/(16\pi^2)=4.6\times10^{-3}$. The logarithmic intervals are $\ln(M_2/M_1)=4.61$ and $\ln(M_1/v)=20.4$. A contribution with an order-one anomalous-dimension coefficient and coupling squared of order unity would therefore scale as $\ln(M_1/v)/(16\pi^2)\simeq0.13$ over the latter interval. This power counting indicates the possible importance of running; the coefficient, sign and cancellation in any relation require its own matched evolution. In particular, the tree-level near-saturation reported below is not a radiative precision result.

\section{Reheating, production spectra and scalar moments}\label{sec:reheat}
The matching bound becomes a production constraint once the injection history and entropy dilution are specified. We use perturbative reheating to relate the inclusive source to the baryon yield and to calculate the scalar populations produced by heavy decays and by the mixed operator. Their distinct momentum scales are essential: primary particles inherit the heavy-decay energy, whereas bath scattering creates much softer particles. Energy density weights a relativistic distribution by $p$, while the scalar variance $I_S$ weights it by $1/(2p)$. For a narrow population, $I_S\simeq\rho_S/(2p^2)$, so a component carrying little energy can dominate the variance and hence the scalar mass correction~\cite{Mukaida:2015density}. The spectra below determine this separation for the same portal constrained by the mixed source.

\subsection{Production and visible entropy}
We specify a perturbative inflaton-dominated era with $2M_1<m_\phi<2M_2$. The inflaton branching fraction into the lighter heavy pair is $B_X$; its remaining energy initially enters visible radiation. The total width defines $T_{\rm RH}$ through $\Gamma_\phi=\sqrt{\pi^2g_*/90}\,T_{\rm RH}^2/\Mpl$, with $g_*=g_{*s}=106.75$ during production. We initialize at $H_i=10^5\Gamma_\phi$ with zero radiation and negligible particle abundances. To isolate the variance generated by these two production channels, we impose $\langle S\rangle=0$ and negligible preexisting scalar occupation. At a common reference epoch $a_*$, the scalar variance defined in Eq.~\eqref{eq:variance} must satisfy
\begin{equation}
 I_S^{\rm preexisting}(a_*)\ll I_S^{\rm prim}(a_*)+I_S^{LH}(a_*).
 \label{eq:initialvariance}
\end{equation}
This is a boundary condition on the inflationary history. Light spectator fields can acquire primordial fluctuations~\cite{Enqvist:2012spectator,Enqvist:2014singlet}; a coherent component would require separate field evolution and would change the assumed light-particle mass structure.

For an inflaton pair coupling proportional to $G_X=\id_2/\sqrt2$, tracing over one member of the pair gives $\rho_X=G_XG_X^\dagger/\Tr(G_XG_X^\dagger)=\id_2/2$. The diagonal decay blocks are proportional to the identity, so this ensemble remains proportional to the identity during prompt decay. This production prescription realizes the equally populated ensemble used in Eq.~\eqref{eq:source}. The hierarchy between $M_1$ and $M_2$ controls the nonresonant interference.

The heavy hidden branching fraction is $b_D=\eta/(1+\eta)$. Prompt decay partitions the inflaton energy into visible and hidden radiation with fixed fractions $f_{\rm vis}=1-B_Xb_D$ and $B_Xb_D$. Writing $\rho_{\rm rad}$ for total radiation and $\rho_{\rm vis}$ for its visible component, the background satisfies
\begin{equation}
 \begin{aligned}
 \dot{\rho}_\phi+3H\rho_\phi&=-\Gamma_\phi\rho_\phi,
 &\dot{\rho}_{\rm rad}+4H\rho_{\rm rad}&=\Gamma_\phi\rho_\phi,\\
 H^2&=\frac{\rho_\phi+\rho_{\rm rad}}{3\Mpl^2},
 &\rho_{\rm vis}&=f_{\rm vis}\rho_{\rm rad} .
 \end{aligned}
 \label{eq:background}
\end{equation}
The visible temperature is determined by $\rho_{\rm vis}$.
The positive scalar interaction $\kappa_{\phi S}\phi^2S^\dagger S$ is chosen with $\kappa_{\phi S}=10^{-20}$. Its contribution to the scalar mass squared reaches $0.447\,\GeV^2$ at the specified initial condensate amplitude and remains negligible in the heavy two-body kinematics. The associated resonance parameter is $1.14\times10^{-24}$. At this time $m_{S,\phi}/H_i=1.90\times10^{-3}$, so the chosen portal provides no heavy-spectator suppression of primordial fluctuations. Its positive sign and the renormalized coefficient of $\phi S^\dagger S$ are matching choices. Appendix~\ref{app:regime} compares them with the induced heavy-loop curvature.

The final visible entropy is needed to normalize both the baryon yield and the scalar spectra. Define $\rho_0=3\Mpl^2\Gamma_\phi^2$, $\Phi=\rho_\phi a^3/\rho_0$,
$R=\rho_{\rm rad}a^4/\rho_0$ and $h=H/\Gamma_\phi$. Then
$\Phi'=-\Phi/h$ and $R'=a\Phi/h$, where the prime denotes differentiation with respect to $\ln a$. The initial values are $\Phi_i=1$, $R_i=0$ and $a_i=(\Gamma_\phi/H_i)^{2/3}$. The asymptotic value $R_\infty$ fixes the common entropy factor
$\mathcal T=(f_{\rm vis}R_\infty)^{-3/4}$. Two heavy particles are produced per relevant inflaton decay, giving
\begin{equation}
 Y_X^{\rm inj}=\frac{3B_XT_{\rm RH}}{2m_\phi}{\cal T},
 \qquad
 Y_B=-C_s\eps_{\rm inc}Y_X^{\rm inj},\qquad C_s=\frac{28}{79}.
 \label{eq:yields}
\end{equation}
The yields use the final visible entropy. The selected CP branch has $\eps_{\rm inc}<0$, so the baryon yield is positive. The conversion factor follows from Standard Model chemical equilibration~\cite{Harvey:1990qw,DOnofrio:2014rug}. Flavour and spectator transport are described in refs.~\cite{Barbieri:1999ma,Abada:2006ea,Nardi:2006fx,Giudice:2003jh,Blanchet:2011density,BhupalDev:2014pfm}; in the prompt weak-washout regime the trace follows \eqr{eq:yields}.

\subsection{Vacuum-width envelope at fixed baryon yield and primary radiation}\label{sec:envelope}
We now ask how small the vacuum widths can be when the same decays must supply a specified baryon yield and primary hidden-energy fraction. Hold fixed the light masses, $r=M_1/M_2$, $\zeta=m_\phi/M_1$, $T_{\rm RH}$, $H_i/\Gamma_\phi$, $Y_B$ and $r_D=B_X\eta/(1+\eta)$. The variables to be minimized over are the inflaton branching fraction $B_X$ and reflection parameter $\xi$; the baryon and radiation conditions then determine $\eta$ and $c_U$. Prompt heavy decay, negligible washout and massless light decay products are retained throughout this minimization.

The source normalization required by Eq.~\eqref{eq:yields}, including visible-entropy dilution, is
\begin{equation}
 \mathcal K(r_D)=\frac{32\pi\zeta v^2(1-r^2)Y_B}
 {3C_s m_-T_{\rm RH}\mathcal T(r_D)},\qquad
 \mathcal T(r_D)=[(1-r_D)R_\infty]^{-3/4}.
 \label{eq:Kd}
\end{equation}
Since the total prompt background is independent of $r_D$, $\mathcal K(r_D)=\mathcal K_0(1-r_D)^{3/4}$ with constant $\mathcal K_0$. The relations $r_D=B_X\eta/(1+\eta)$ and
$B_X=\mathcal K(r_D)(1+\eta)/(\xi c_U)$ give
\begin{equation}
 \eta=\frac{r_D}{B_X-r_D},\qquad
 c_U=\frac{\mathcal K(r_D)}{\xi(B_X-r_D)},\qquad
 0<r_D<B_X\le1,\quad0<\xi\le1.
 \label{eq:portalelimination}
\end{equation}
The weighted sum $\mathcal W_\nu=\sum_{i=2,3}\Gamma_i^{(0)}/m_i$ removes the redistribution of coupling strength between the two massive rows: the opposite interference terms in Eq.~\eqref{eq:reflectionwidths} cancel. Substituting the eliminated portal parameters gives
\begin{equation}
 \mathcal W_\nu=
 \frac{r_D m_+^2+\mathcal K(r_D)^2m_-^2/(\xi^2r_D)}{16\pi v^2(B_X-r_D)}
 \ge\frac{r_D m_+^2+\mathcal K(r_D)^2m_-^2/r_D}{16\pi v^2(1-r_D)}
 \equiv\mathcal W_{\nu,\min}(r_D).
 \label{eq:radiationenvelope}
\end{equation}
Equality requires $B_X=\xi=1$, with
$\eta=r_D/(1-r_D)$ and $c_U=\mathcal K(r_D)/(1-r_D)$. It is an infimum if
$B_X<1$ is imposed strictly. Perturbativity, prompt decay and weak washout
can remove the equality point from the physical domain while preserving
the inequality.

The individual widths retain the interference that cancels in this sum. At fixed $r_D$, write $t=1/\xi\ge1$ and suppress the argument of $\mathcal K(r_D)$. The numerators are
$N_{3,2}(t)=r_D m_+^2+\mathcal K^2m_-^2t^2/r_D\pm2\mathcal K m_+ m_-(t-1)$.
The heavier-row numerator increases with $t$, while the lighter-row numerator has its unconstrained minimum at $t=r_D m_+/(\mathcal K m_-)$. Taking the allowed minimum in $t$ and the largest denominator, $B_X=1$, gives
\begin{equation}
 \begin{aligned}
 \Gamma_3^{(0)}&\ge\frac{m_3}{32\pi v^2(1-r_D)}
 \left(r_D m_+^2+\frac{\mathcal K^2m_-^2}{r_D}\right),\\
 \Gamma_2^{(0)}&\ge\frac{m_2}{32\pi v^2(1-r_D)}
 \begin{cases}
 r_D m_+^2+\mathcal K^2m_-^2/r_D,&r_D\le \mathcal K m_-/m_+,\\
 2\mathcal K m_+ m_-,&r_D\ge \mathcal K m_-/m_+.
 \end{cases}
 \end{aligned}
 \label{eq:radiationindividual}
\end{equation}
For $\nu_2$ the second branch is reached at $\xi=\mathcal K m_-/(r_D m_+)$.
These relations concern the primary reheating fraction $r_D$ and massless
decay products. The later interacting radiation abundance requires the
coupled evolution discussed in Section~\ref{sec:medium}.

Separate redshifting of the primary component through visible entropy release defines the reference radiation normalization
\begin{equation}
 \Delta N_{\rm eff}^{\rm prim}=
 \frac47\frac{r_D}{1-r_D}\,g_*(T_{\rm RH})
 \left[\frac{g_{*s}(T_{\nu{\rm dec}})}
             {g_{*s}(T_{\rm RH})}\right]^{4/3},
 \qquad g_{*s}(T_{\nu{\rm dec}})=10.75 .
 \label{eq:neff}
\end{equation}
Subsequent neutrino decay transfers energy between active and hidden particles in a common interacting system. Eq.~\eqref{eq:neff} labels the initial radiation component.

\subsection{Primary spectrum and inverse-energy moment}
The primary spectrum determines how the heavy-decay energy is spread by the boost and by production at different scale factors. Both effects matter for the inverse-energy weight. The heavy particles have speed $\beta_X=\sqrt{1-4M_1^2/m_\phi^2}$. For an unpolarized pair, each massless hidden decay product has a uniform energy distribution between $E_-=m_\phi(1-\beta_X)/4$ and $E_+=m_\phi(1+\beta_X)/4$. With comoving momentum $q_{\rm c}=ap$ and normalized decay probability $\dd P=-\dd\Phi$, the number spectrum is
\begin{equation}
 \frac{\dd P_D}{\dd\ln q_{\rm c}}=
 \int\dd\ln a\,\frac{\Phi(a)}{h(a)}
 \frac{q_{\rm c}}{a(E_+-E_-)}
 \Theta(q_{\rm c}-aE_-)\Theta(aE_+-q_{\rm c}).
 \label{eq:spectrum}
\end{equation}
Each hidden decay creates one scalar and one Weyl fermion; these species have the same primary momentum spectrum. Their charge-conjugate populations are included in the number densities.

Conservation of the hidden charge relates the particle asymmetries but
does not set either one to zero. With the same particle-minus-antiparticle
convention as $\eps_{\rm inc}$, the mixed hidden source contains
$\Imn\Tr(D_{12}K_{12}^T)=J_+$ and
$\Imn\Tr(D_{12}K_{12}^\dagger)=-J_-=0$.
At the endpoint the pure-hidden contraction also has zero imaginary
trace, since $D_{12}D_{12}^T=c_U^2\kappa_1\kappa_2\sigma_1$.
The number asymmetry per total heavy decay is therefore
$\eps_n=\eps_{\rm inc}$. Each hidden decay produces one $n$ and one
$S$, so $\Delta N_n=\Delta N_S$ and their opposite charges give
$\Delta Q_D=0$. Writing $n_{a,+}$ and $n_{a,-}$ for particle and antiparticle densities,
with $n_S=n_{S,+}+n_{S,-}$, the primary fractional number asymmetries are
\begin{equation}
 \frac{n_{n,+}-n_{n,-}}{n_{n,+}+n_{n,-}}
 =\frac{n_{S,+}-n_{S,-}}{n_{S,+}+n_{S,-}}
 =\frac{\eps_{\rm inc}}{b_D}\simeq-1.84\times10^{-4}.
 \label{eq:primarycharge}
\end{equation}
The charge-symmetric approximation thus retains the CP-even primary moments
to the order used here. Charge-resolved distributions remain the appropriate
variables for subsequent coupled evolution.

Earlier injection redshifts further and therefore receives greater weight in the variance than in the energy density. Correspondingly, the mean comoving momentum depends on $\langle a\rangle=R_\infty$, whereas its inverse moment depends on $\langle a^{-1}\rangle=\int\dd\ln a\,\Phi/(ah)$. After reheating,
\begin{equation}
 \frac{\langle p_D\rangle}{T}=\frac{m_\phi R_\infty}{4aT},
 \qquad
 T\langle p_D^{-1}\rangle=
 \frac{aT\,\langle a^{-1}\rangle}{E_+-E_-}
 \ln\frac{E_+}{E_-},
 \qquad
 aT=T_{\rm RH}(f_{\rm vis}R_\infty)^{1/4}.
 \label{eq:moments}
\end{equation}
For collisionless primary particles, visible entropy release multiplies momenta relative to the neutrino temperature by $(10.75/106.75)^{1/3}$. The inverse moment fixes the density part of the scalar variance,
\begin{equation}
 I_S=\langle S^\dagger S\rangle_{\rm density}
   =\int\frac{\dd^3p}{(2\pi)^3}
          \frac{f_S(p)+f_{S^\dagger}(p)}{2\omega_p}
   =\frac{n_S}{2}\langle\omega^{-1}\rangle .
 \label{eq:variance}
\end{equation}
The factor $1/2$ follows from the complex-scalar normalization, with $n_S$ including both charges. Retaining the inverse moment avoids replacing a broad spectrum by a single characteristic energy.

\subsection{Production from the visible bath}\label{sec:bath}
The bath contribution tests whether the primary spectrum also controls the variance. The mixed operator produces scalars through $LH\to\bar n S^\dagger$ and its charge conjugate, as in mixed-operator freeze-in~\cite{Blazek:2025effective}. We evaluate this contribution with massless Fermi--Dirac leptons, Bose--Einstein Higgs particles and dilute outgoing states at Born order. The inclusive cross section is $\norm{C_{Ln}}^2/(4\pi)$, including the $\sqrt2$ portal factor, weak components and charge conjugation. The chiral amplitude gives the scalar angular distribution $(1+c_*)/2$ in the centre-of-mass frame, where $c_*$ is the cosine relative to the incoming lepton. This angular dependence determines how the incident energy is divided between the final-state particles.

Writing $\mathcal C^2=\norm{C_{Ln}}^2$, the number, scalar-energy and variance sources are
\begin{equation}
 \begin{aligned}
 \dot n_S^{LH}+3Hn_S^{LH}
   &=Q_n=\frac{3\zeta(3)^2}{16\pi^5}\mathcal C^2T^6,\\
 \dot{\rho}_S^{LH}+4H\rho_S^{LH}
   &=Q_{\rho_S}=\frac{\zeta(3)}{144\pi}\mathcal C^2T^7,\\
 \dot{I}_S^{LH}+2HI_S^{LH}
   &=Q_I=c_I\mathcal C^2T^5,\qquad c_I=2.206\times10^{-4}.
 \end{aligned}
 \label{eq:bathmoments}
\end{equation}
The combined scalar and fermion energy source is
$Q_D=13\zeta(3)\mathcal C^2T^7/(960\pi)$. The scalar receives $20/39$ of this energy and the fermion receives $19/39$. The unequal partition follows from the chiral angular distribution and the different incoming statistics. Appendix~\ref{app:bathspectrum} derives these coefficients and the differential scalar production spectrum $\mathcal S(\ell)$, with $\ell=p_S/T$ at production.

To compare the two populations after visible entropy release, define the conserved reference temperature normalization
$\mathcal Q_\nu=aT_\nu=T_{\rm RH}(f_{\rm vis}R_\infty)^{1/4}(106.75/10.75)^{1/3}$.
Collisionless redshifting gives
\begin{equation}
 \begin{aligned}
 \mathcal N_{LH}(y)&\equiv\frac1{T_\nu^3}
       \frac{\dd n_S^{LH}}{\dd\ln y}
  =\frac{\mathcal C^2}{16\pi^5\mathcal Q_\nu^3}
       \int\dd\ln a\,\frac{a^3T^6}{H}
           \mathcal S\left(\frac{y\mathcal Q_\nu}{aT}\right),
       \qquad y=\frac p{T_\nu},\\
 \frac{I_S^{LH}}{T_\nu^2}&=\mathfrak K_I\mathcal C^2,\\
 \mathfrak K_I&=\frac{c_I}{\mathcal Q_\nu^2}
       \int\dd\ln a\,\frac{a^2T^5}{H}.
 \end{aligned}
 \label{eq:bathspectrum}
\end{equation}
The primary spectrum is
$\mathcal N_{\rm prim}(y)=(n_S^{\rm prim}/T_\nu^3)\dd P_D/\dd\ln y$.
Their sum defines the production reference used below: each component is redshifted independently, with subsequent neutrino conversion omitted. A distribution-level treatment retains the different momentum weights needed for energy and variance~\cite{Du:2021distribution}.

During the matter-dominated reheating interval, $T\propto a^{-3/8}$ and $H\propto a^{-3/2}$. The comoving source of the momentum moment weighted by $p^j$ then scales as $a^{(18+5j)/8}$. The variance, number and energy sources all grow toward the end of reheating despite the earlier temperature maximum. Their radiation-era integrands subsequently scale as $a^{-1}$; the remaining contribution is integrated analytically and included in \eqr{eq:bathspectrum}. The constant relativistic degrees of freedom and massless bath specify the production approximation. Energy transfer enters the visible and hidden equations with terms $-Q_D$ and $+Q_D$, respectively; its small backreaction is governed by the transferred energy fraction.

\subsection{The mixed source and the scalar variance}\label{sec:variancebound}
A nonzero mixed source restricts how much the bath-produced variance can be reduced by threshold interference. For a specified background, $\mathfrak K_I>0$ is fixed by the production integral and has mass dimension two. Its product with the squared portal norm is therefore a positive, dimensionless scalar moment. At fixed threshold norms, Eqs.~\eqref{eq:domain} and \eqref{eq:bathspectrum} give
\begin{equation}
 \frac{I_S^{LH}}{T_\nu^2}\ge
 \mathfrak K_I\left[a_1^2+a_2^2
             -2\sqrt{a_1^2a_2^2-j^2}\right].
 \label{eq:variancebound}
\end{equation}
Here $a_I$ and $j$ are the matching quantities defined above \eqr{eq:domain}. With an independent $\Delta C$, the bracket is replaced by the square of the right-hand side of \eqr{eq:thresholdbound}. This bounds the variance produced through the open mixed channel in the stated dilute production regime. Its scale depends on the portal strengths and the reheating history. The normalized-source form in Eq.~\eqref{eq:normalizedbound} gives the complementary result
\begin{equation}
 \frac{I_S^{LH}}{T_\nu^2}\ge\mathfrak K_I
 \left[\frac{16\pi(1-r^2)}{M_1}|\eps_+|\right]^2,
 \qquad
 \frac{I_S^{LH}}{T_\nu^2}=\frac{8\pi\mathfrak K_I}{v^2}\mathcal W_\nu .
 \label{eq:sourcevariance}
\end{equation}
The second equality uses the massive-row support condition in Eq.~\eqref{eq:weightedwidth}; it holds throughout the reflection family. These relations constrain the production reference. Subsequent inverse reactions and self-consistent dispersion change the collision evolution and require recomputing the physical variance.

For a massless component, define $p_{\rm eff}^2=\rho_S/(2I_S)=\langle p\rangle/\langle p^{-1}\rangle$. The relative contributions of bath and primary particles obey
\begin{equation}
 \frac{I_S^{LH}}{I_S^{\rm prim}}
 =\frac{\rho_S^{LH}}{\rho_S^{\rm prim}}
       \left(\frac{p_{\rm eff,prim}}{p_{\rm eff,LH}}\right)^2 .
 \label{eq:momentratio}
\end{equation}
The momentum ratio supplies the information needed to compare the two variances at a specified energy fraction. The benchmark below evaluates both factors from the production spectra.

\section{Benchmark correlations and production moments}\label{sec:numbers}
\subsection{The source and the low-energy coupling}\label{sec:benchmark}
The endpoint texture connects a nonzero inclusive mixed source to equal-strength decay couplings for the two massive neutrinos. We evaluate this connection at the central parameters in Table~\ref{tab:inputs}, then use the same parameters to compare the primary and bath-produced scalar moments. We use the normal-ordering NuFIT 6.0 IC24 inputs including Super-Kamiokande atmospheric data~\cite{Esteban:2024eli}, with $m_1=0$, $\sin^2\theta_{12}=0.308$, $\sin^2\theta_{23}=0.470$, $\sin^2\theta_{13}=0.02215$, $\delta_{\rm CP}=212^\circ$ and zero Majorana phases. The resulting masses are $m_2=8.65\times10^{-3}\,\eV$ and $m_3=5.01\times10^{-2}\,\eV$. The nominal baryon target $Y_B=8.60\times10^{-11}$ fixes $B_X$ through \eqr{eq:yields}, using the cosmological normalization in refs.~\cite{Planck:2018vyg,ParticleDataGroup:2024cfk}.

\begin{table}[htbp]
 \centering

 \begin{tabular}{lll}
 \toprule
 Quantity & Value & Role\\
 \midrule
 $M_1$ & $1.25325\times10^{11}\,\GeV$ & lighter heavy level\\
 $M_2/M_1$ & $100$ & heavy hierarchy\\
 $m_\phi/M_1$ & $2.5$ & inflaton decay kinematics\\
 $T_{\rm RH}$ & $5.00\times10^7\,\GeV$ & inflaton width\\
 $H_i/\Gamma_\phi$ & $10^5$ & perturbative-era initial state\\
 $\Delta m_{21}^2$ & $7.49\times10^{-5}\,\eV^2$ & oscillation input\\
 $\Delta m_{31}^2$ & $2.513\times10^{-3}\,\eV^2$ & oscillation input\\
 $\eta$ & $0.028687$ & hidden-to-visible width ratio\\
 $c_U$ & $1.54296$ & portal normalization\\
 $m_{S,0}$ & $0$ & massless scalar benchmark\\
 $\lambda_{HS}$ & $0$ & Higgs-portal matching\\
 $\kappa_{\phi S}$ & $10^{-20}$ & inflaton--scalar interaction\\
 $\lambda_S$ & independent, $\lambda_S\ge0$ & density response in Section~\ref{sec:medium}\\
 \bottomrule
 \end{tabular}
 \caption{Central inputs and scalar matching parameters in the leading-order convention of Section~\ref{sec:running}. The portal parameters select a representative perturbative point with open massless decay channels; they are independent inputs. Displayed digits identify the calculation, and derived values are quoted to three significant figures. Independent higher-threshold contributions to the three coefficients in \eqr{eq:matching} are set to zero.}
 \label{tab:inputs}
\end{table}

The seesaw fixes $\kappa_1=1.22\times10^{-4}$ and $\kappa_2=8.58\times10^{-3}$.
The hidden block is $D_{22}=0.712\id_2$, with $\Gamma_2^{\rm tot}/M_2=0.0287$.
At the lighter level $\Gamma_1^{\rm tot}=6.24\times10^5\,\GeV$ and $b_D=0.0279$.
The two heavy levels are well separated on the scale of their widths. The mixed invariant is $J_+=3.22\times10^{-6}$, and the inclusive asymmetry is $\eps_{\rm inc}=-5.12\times10^{-6}$. The source matrix $\eps$ has nonzero eigenvalues $-2.59\times10^{-6}$ and $-2.54\times10^{-6}$.

The mixed coefficient is $b=(1.64\times10^{-16}-6.24\times10^{-15}i)\,\GeV^{-1}$.
Its norm is $\norm{C_{Ln}}=8.83\times10^{-15}\,\GeV^{-1}$; the lower bound in \eqr{eq:normbound} lies a relative $3.5\times10^{-4}$ below this value. This near-saturation follows from the dominance of the heavier-level contribution. The resulting coupling is $g_\nu=1.54\times10^{-12}$, giving the vacuum lifetimes in Table~\ref{tab:outputs}. We use $\hbar=6.582119569\times10^{-25}\,\GeV\,\mathrm{s}$ for conversion to seconds.

\subsection{Vacuum lifetimes and scalar thresholds}\label{sec:lifetime}
\begin{figure}[htbp]
 \centering
 \begin{subfigure}{0.49\linewidth}
  \includegraphics[width=\linewidth]{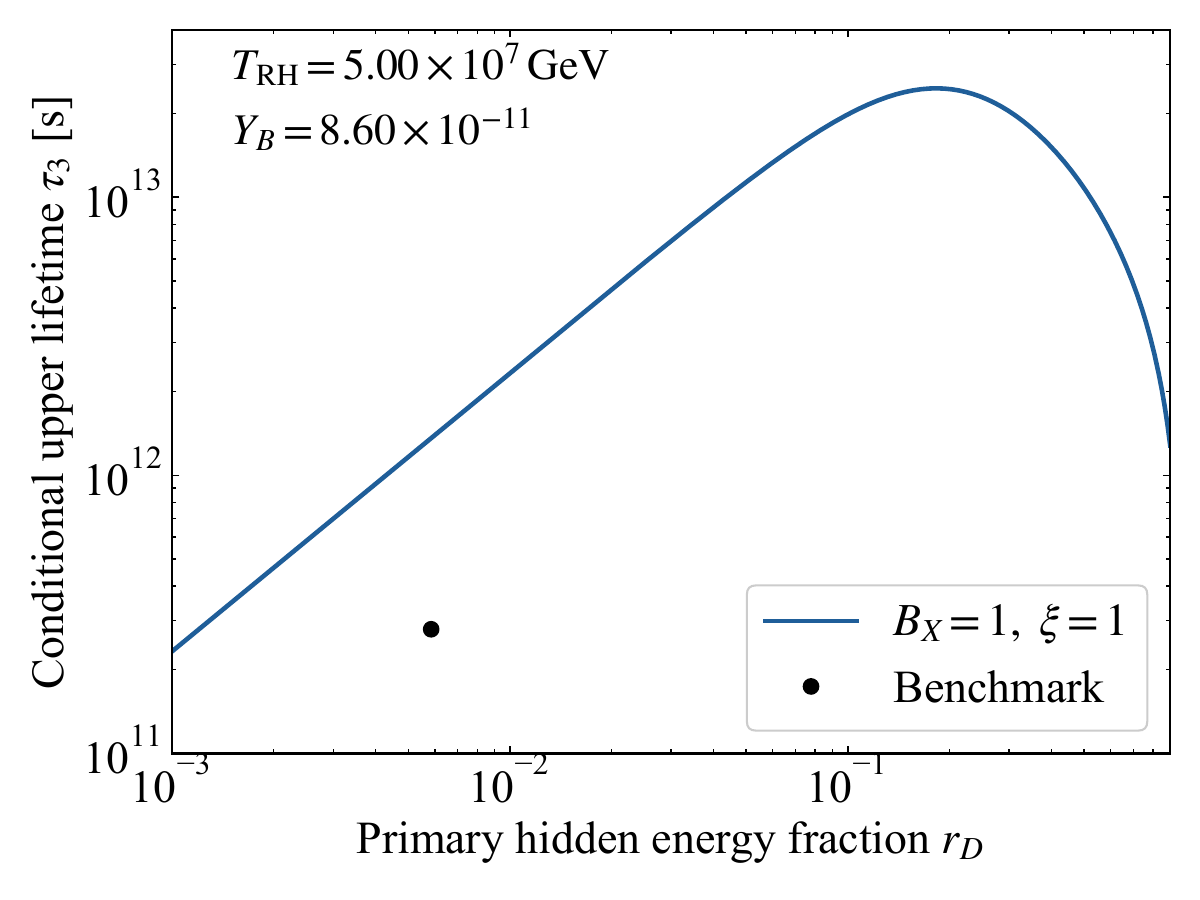}
  \caption{Baryon-normalized lifetime envelope.}\label{fig:radlife}
 \end{subfigure}\hfill
 \begin{subfigure}{0.49\linewidth}
  \includegraphics[width=\linewidth]{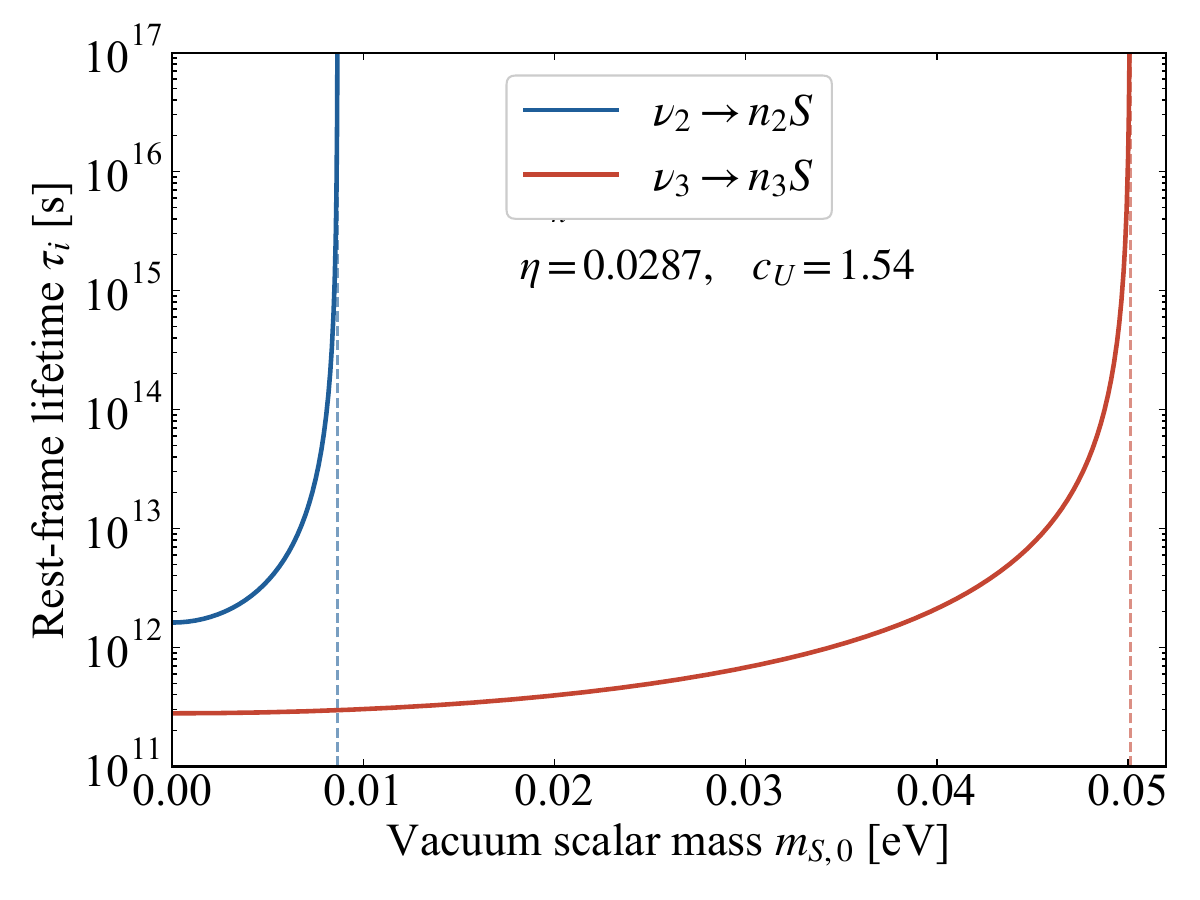}
  \caption{Vacuum scalar-mass thresholds.}\label{fig:thresholds}
 \end{subfigure}
 \caption{Vacuum neutrino lifetimes from the mixed operator. (a) Algebraic upper envelope for $\tau_3$ at $T_{\rm RH}=5.00\times10^7\,\GeV$ and $Y_B=8.60\times10^{-11}$, with fixed mass ratios. The boundary has $B_X=\xi=1$; the marker denotes the central point. Further dynamical and perturbative restrictions can reduce attainability. The horizontal coordinate labels the independently redshifted primary radiation component. (b) The vacuum scalar mass is varied at the central couplings with massless hidden fermions. Vertical lines locate $m_{S,0}=m_i$.}
 \label{fig:lifetimes}
\end{figure}

Figure~\ref{fig:lifetimes} separates the algebraic width envelope from the effect of a scalar threshold. Panel~\ref{fig:radlife} shows the upper vacuum lifetime of $\nu_3$ following from Eq.~\eqref{eq:radiationindividual}, with $\mathcal K_0=0.316$. At the benchmark radiation fraction $r_D=5.85\times10^{-3}$, the ceiling is $1.36\times10^{12}\,\mathrm{s}$, above the central lifetime because the benchmark has $B_X<1$. The algebraic ceiling peaks at $2.46\times10^{13}\,\mathrm{s}$ near $r_D=0.183$. Equality requires $B_X=\xi=1$ before imposing perturbativity, prompt decay and weak washout. The fixed-$c_U$ and fixed-$\alpha$ minima in Eqs.~\eqref{eq:fixedcu} and \eqref{eq:fixedalpha} describe more restricted trajectories inside this domain.

Panel~\ref{fig:thresholds} holds the central couplings fixed and varies the vacuum scalar mass. The two thresholds separate a region with both neutrino decays, a region with only the heavier-neutrino decay, and a region with closed two-body channels. The rapid lifetime growth near threshold follows the squared phase-space factor in \eqr{eq:scalarwidth}. A scalar mass that is negligible for heavy-particle production can therefore determine which light-neutrino channel is active.

The vacuum lifetime is a rest-frame property. In a relativistic cosmological ensemble, time dilation, inverse reactions and the daughter distributions enter conversion, while angular damping probes the perturbed collision operator. Its relativistic massless-daughter scaling can differ parametrically from the total decay rate~\cite{Barenboim:2020decay}; the active-daughter mass effects studied in ref.~\cite{Chen:2022lifetime} require that paper's distinct spectrum. Comparing $\tau_i$ with the recombination age alone therefore gives no free-streaming exclusion. The numerical point here is a matching and production benchmark. Its cosmological viability with the nonthermal scalar state requires homogeneous and perturbed evolution including Bose enhancement.

\subsection{Reheating normalization and scalar moments}
At the central parameter point, primary scalars dominate the energy density and the softer bath scalars dominate the variance. This comparison uses prompt heavy decay, dilute outgoing states and a common final visible-entropy normalization; the following scale comparisons test those conditions.

The reheating solution gives $\Gamma_\phi=3.51\times10^{-3}\,\GeV$ and $H_i=351\,\GeV$.
The initial ratio $H_i/\Gamma_1^{\rm tot}=5.63\times10^{-4}$ decreases during the main production epoch. The maximum visible temperature is $5.42\times10^8\,\GeV$, giving $M_1/T_{\max}=231$. The injection boost is $m_\phi/(2M_1)=1.25$, so the initial ratio to the time-dilated heavy width remains $7.04\times10^{-4}$. These hierarchies support prompt decay in \eqr{eq:background}. Thermal inverse production is exponentially suppressed~\cite{HahnWoernle:2009full,Nardi:2007scatter,Fong:2010gauge,Racker:2018nlo}. For a hard particle of energy $E_h$, reaching a heavy pole with a bath particle of energy $k$ requires $4E_hk\ge M_1^2$. The primary endpoint $E_h=M_1$ requires $k/T_{\max}\ge57.8$; even a direct visible inflaton daughter with $E_h=m_\phi/2$ requires $k/T_{\max}\ge46.2$. These hard--thermal routes remain suppressed. We assume rapid thermalization of visible inflaton products; hard--hard reactions before thermalization would depend on their separate density and lifetime. The time-integrated Weinberg-operator washout is $6.54\times10^{-6}$. Table~\ref{tab:regime} in Appendix~\ref{app:regime} gives the time-integrated hidden reaction rates per particle and occupation numbers supporting the dilute production regime.

Table~\ref{tab:outputs} gives the source, vacuum lifetimes and reheating normalization. The mixed-operator energy transfer contributes $2.74\times10^{-7}$ to the total radiation fraction, or $4.69\times10^{-5}$ of the primary hidden energy. Its effect on the energy balance is below the displayed precision of the yields normalized to the baryon abundance.

\begin{table}[htbp]
 \centering

 \begin{tabular}{ll@{\qquad}ll}
 \toprule
 Quantity & Value & Quantity & Value\\
 \midrule
 $\eps_{\rm inc}$ & $-5.12\times10^{-6}$ &
 $g_\nu$ & $1.54\times10^{-12}$\\
 $\tau_2$ & $1.62\times10^{12}\,\mathrm{s}$ &
 $\tau_3$ & $2.80\times10^{11}\,\mathrm{s}$\\
 $B_X$ & $0.210$ &
 $Y_X^{\rm inj}$ & $4.74\times10^{-5}$\\
 $R_\infty$ & $1.09$ &
 $\mathcal T$ & $0.943$\\
 $Y_B$ & $8.60\times10^{-11}$ &
 $r_D$ & $5.85\times10^{-3}$\\
 $\Delta N_{\rm eff}^{\rm prim}$ & $0.0168$ &
 $\mathcal Q_\nu$ & $1.10\times10^8\,\GeV$\\
 \bottomrule
 \end{tabular}
 \caption{Source, vacuum-decay and reheating results at the parameter point in Table~\ref{tab:inputs}. The baryon yield fixes $B_X$. The hidden radiation quantities refer to the primary component redshifted independently.}
 \label{tab:outputs}
\end{table}

The primary spectrum produced by sequential inflaton and heavy-particle decays has $\beta_X=0.6$, $\langle a^{-1}\rangle=2.04$ and
$\langle p_D\rangle/T=1.67\times10^3$ after reheating. Visible entropy release gives $\langle p_D\rangle/T_\nu=777$. Table~\ref{tab:moments} separates the primary and bath-produced scalar moments at the common reference temperature. Replacing $\langle p^{-1}\rangle$ by $\langle p\rangle^{-1}$ underestimates the primary variance by a factor of $2.57$. The bath-produced scalars have a much lower mean momentum, $\langle p_S\rangle/T_\nu=1.27$, and a different spectral shape.

\begin{table}[htbp]
 \centering

 \begin{tabular}{lll}
 \toprule
 Quantity & Primary heavy decays & Bath production\\
 \midrule
 $n_S/T_\nu^3$ & $6.23\times10^{-6}$ & $1.84\times10^{-7}$\\
 $\rho_S/T_\nu^4$ & $4.84\times10^{-3}$ & $2.33\times10^{-7}$\\
 $I_S/T_\nu^2$ & $1.03\times10^{-8}$ & $1.14\times10^{-7}$\\
 $\langle p_S\rangle/T_\nu$ & $777$ & $1.27$\\
 $T_\nu\langle p_S^{-1}\rangle$ & $3.30\times10^{-3}$ & $1.24$\\
 $\langle p_S\rangle\langle p_S^{-1}\rangle$ & $2.57$ & $1.57$\\
 \bottomrule
 \end{tabular}
 \caption{Scalar moments after visible entropy release, with both production components redshifted collisionlessly. Bath production uses the massless Born approximation with Fermi--Dirac leptons, Bose--Einstein Higgs particles and dilute outgoing states. Number densities include both charges; energy entries contain scalars only. The total reference variance is $I_S^{\rm ref}/T_\nu^2=1.24\times10^{-7}$.}
 \label{tab:moments}
\end{table}
\FloatBarrier

\begin{figure}[htbp]
 \centering
 \begin{subfigure}{0.49\linewidth}
  \includegraphics[width=\linewidth]{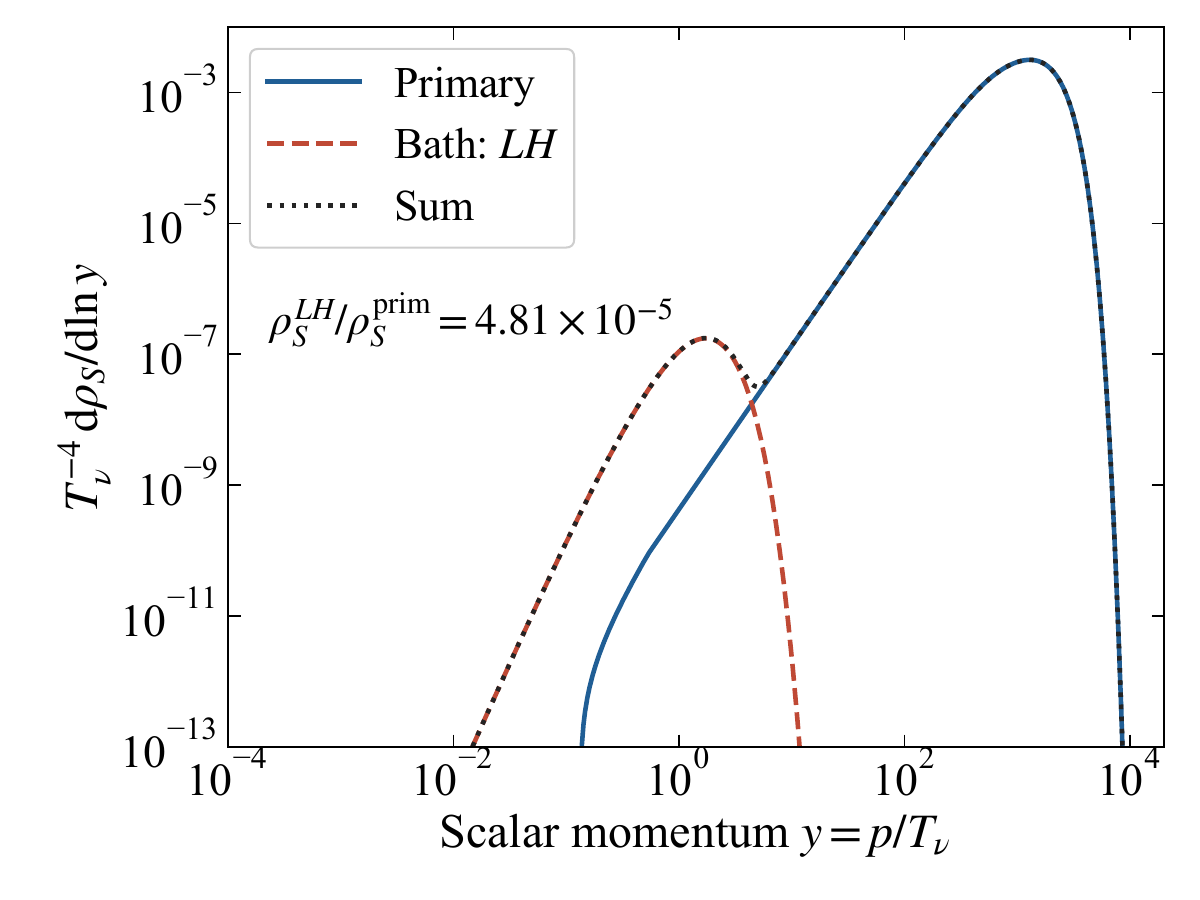}
  \caption{Scalar energy density.}\label{fig:energyspectrum}
 \end{subfigure}\hfill
 \begin{subfigure}{0.49\linewidth}
  \includegraphics[width=\linewidth]{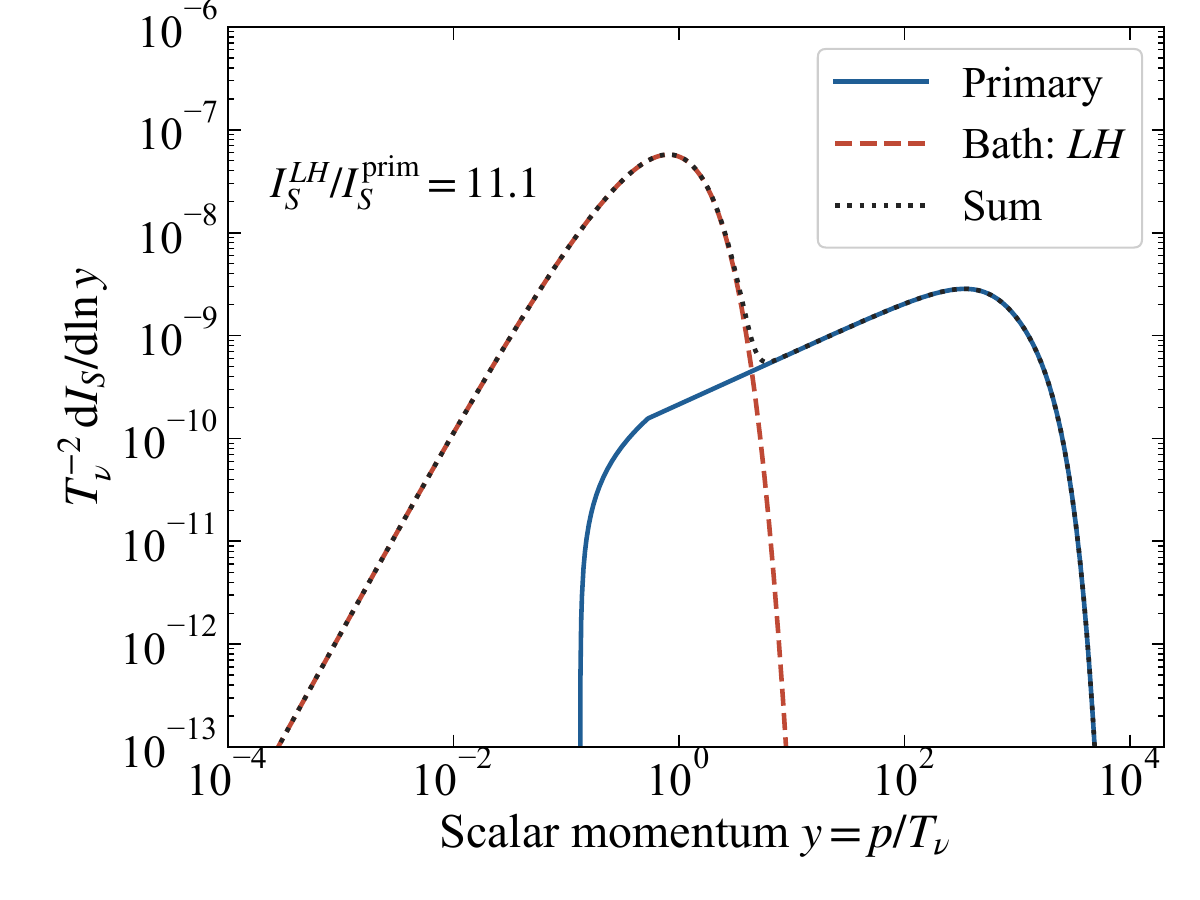}
  \caption{Scalar variance.}\label{fig:variancespectrum}
 \end{subfigure}
 \caption{Hard primary particles dominate scalar energy and softer bath particles dominate scalar variance. The spectra are evaluated after visible entropy release, with $y=p/T_\nu$ and $\mathcal N_c(y)=T_\nu^{-3}\dd n_S^c/\dd\ln y$. Panel~\ref{fig:energyspectrum} shows $y\mathcal N_c(y)$, which integrates to $\rho_S^c/T_\nu^4$. Panel~\ref{fig:variancespectrum} shows $\mathcal N_c(y)/(2y)$, which integrates to $I_S^c/T_\nu^2$. Each panel includes the sum. The bath calculation uses massless Fermi--Dirac leptons, Bose--Einstein Higgs particles and dilute outgoing states at Born order.}
 \label{fig:spectrum}
\end{figure}

Figure~\ref{fig:spectrum} displays the contributions to the scalar energy density and variance per logarithmic momentum interval. Bath production supplies $2.95\%$ of the primary scalar number and $4.81\times10^{-5}$ of its scalar energy, yet its variance is larger by a factor of $11.1$. The bath therefore carries $91.7\%$ of the combined reference variance. The effective momenta in \eqr{eq:momentratio} are $p_{\rm eff,prim}/T_\nu=485$ and $p_{\rm eff,LH}/T_\nu=1.01$, explaining this reversal. The scalar energy ratio differs slightly from the total hidden energy ratio because the bath partitions energy according to $20:19$, while primary decays give equal scalar and fermion energies.

\FloatBarrier
\Needspace{3\baselineskip}
Replacing the incident distributions by Maxwell--Boltzmann forms gives $c_I^{\rm MB}=1/(16\pi^5)$ and $I_S^{LH}/I_S^{\rm prim}=10.24$ on the same background. Bath dominance thus persists in this statistics comparison. Varying the initial epoch from $H_i/\Gamma_\phi=10^4$ to $10^6$, with other inputs fixed and $B_X$ refitted, changes the ratio from $11.33$ to $10.93$. Plasma masses and additional thermal cuts remain a separate physical correction to the massless Born approximation. For the benchmark reheating history, $\mathfrak K_I=1.46\times10^{21}\,\GeV^2$. The threshold contributions are in quadrature, so the central texture saturates the norm domain in \eqr{eq:domain} and the corresponding variance bound in \eqr{eq:variancebound}. This relates the dominant reference variance to the same mixed coefficient that fixes the vacuum neutrino widths.

The division of variance between the two populations also depends on the heavy scale. Hold the light masses, $\eta$, $c_U$, $T_{\rm RH}$, $H_i/\Gamma_\phi$ and the heavy and inflaton mass ratios fixed. The baryon fit leaves $B_X$ and the background unchanged; the mixed coefficient and vacuum widths remain fixed by \eqr{eq:seesaw}. Primary scalar number scales as $M_1^{-1}$ and primary momentum as $M_1$, whereas bath production is unchanged. Hence
$I_S^{LH}/I_S^{\rm prim}=11.1[M_1/(1.25325\times10^{11}\,\GeV)]^2$.
The scalar energy ratio stays constant, and equal variances occur at $M_1=3.77\times10^{10}\,\GeV$. These scalings apply within the prompt, nonresonant and perturbative family. They show how the relative contributions to the scalar variance can change while the vacuum interaction remains fixed.

\section{Scalar response and the domain of kinetic evolution}\label{sec:medium}
The production variance can affect neutrino decay by changing the scalar dispersion, especially in the soft modes with the largest stimulated gain. The independently redshifted production spectra supply the reference scalar variance. To isolate the response to dispersion, the gain calculation prescribes the scalar pole mass and uses fixed Fermi--Dirac parent distributions and empty hidden-fermion states. The quartic comparison estimates this mass from the reference variance with the Yukawa self-energy set to zero. We then derive conserving Hartree relations for self-consistent scalar dispersion at specified occupations. A cosmological transport prediction requires evolving these distributions together. We use the orthogonal endpoint $\xi=1$; for $0<\xi<1$, the overlap in Eq.~\eqref{eq:reflectioncouplings} requires hidden-flavour density matrices or a justified decoherence approximation.

\subsection{Quantum collision terms}
The decay and inverse-decay collision terms specify which statistical factors control local growth and which moments are conserved during conversion. For the central texture in \eqr{eq:texture}, the light interaction connects the two massive neutrinos to orthogonal charged Weyl states and a shared scalar. In the leading charge-symmetric approximation quantified by Eq.~\eqref{eq:primarycharge}, assign two populated helicity or charge states to each of $\nu_i$, $n_i$ and $S$. With parent energy $E=\sqrt{p^2+m_i^2}$ and massless decay products, define
\begin{equation}
 {\cal F}_i(E,k)=
 f_{\nu_i}(p)[1-f_{n_i}(k)][1+f_S(E-k)]
 -[1-f_{\nu_i}(p)]f_{n_i}(k)f_S(E-k).
 \label{eq:statistical}
\end{equation}
The occupation functions are per populated state. Angular integration gives
\begin{equation}
 \begin{aligned}
 (\partial_t-Hp\partial_p)f_{\nu_i}(p)
 &=-\frac{m_i\Gamma_i^{(0)}}{Ep}
       \int_{(E-p)/2}^{(E+p)/2}\dd k\,{\cal F}_i(E,k),\\
 (\partial_t-Hk\partial_k)f_{n_i}(k)
 &=\frac{m_i\Gamma_i^{(0)}}{k^2}
       \int_{k+m_i^2/(4k)}^\infty\dd E\,{\cal F}_i(E,k),\\
 (\partial_t-Hk\partial_k)f_S(k)
 &=\sum_{i=2,3}\frac{m_i\Gamma_i^{(0)}}{k^2}
       \int_{k+m_i^2/(4k)}^\infty\dd E\,{\cal F}_i(E,E-k).
 \end{aligned}
 \label{eq:collision}
\end{equation}
When the decay products have negligible occupation numbers, the parent decay rate reduces to the time-dilated rate $m_i\Gamma_i^{(0)}/E$. Detailed balance holds when $\mu_{\nu_i}^{\rm chem}=\mu_{n_i}^{\rm chem}+\mu_S^{\rm chem}$ in a common thermal state. The number moments of these reactions conserve $a^3(n_{\nu_i}+n_{n_i})$ and $a^3(n_S-n_{n_2}-n_{n_3})$. Event-by-event energy conservation gives
\begin{equation}
 \dot{\rho}_\nu+3H(\rho_\nu+P_\nu)=-Q_{\nu D},\qquad
 \dot{\rho}_D+3H(\rho_D+P_D)=Q_{\nu D}.
 \label{eq:energy}
\end{equation}
The contact operator $C_{nn}$ supplies additional reactions in the full effective theory; the displayed number combinations belong to the decay network.

The microscopic loss rate and the relaxation of a perturbation probe different occupation factors. At fixed distributions of the decay products, the parent relaxation coefficient contains $1-f_n+f_S$. For a parent at rest in a common thermal bath with zero chemical potentials it is $\Gamma^{(0)}\coth[m/(2T)]$, while its loss rate is $\Gamma^{(0)}/[1-e^{-m/T}]$~\cite{Kiessig:2010thermal,PerezGonzalez:2026thermal}. In the present production history the two nonthermal scalar components in Figure~\ref{fig:spectrum} enter the momentum-dependent statistical factor in \eqr{eq:statistical}.

\subsection{The soft scalar source and a finite pole mass}
The scalar variance is sensitive to soft particles, so we need the gain in individual momentum modes as well as its integral over phase space. Stimulated neutrino decay supplies the factor $1+f_S$ in the established decay and inverse-decay framework~\cite{Kaiser:1993lasing,Starkman:1993lasing}. Here $q$ is the physical scalar momentum. We specialize the parent distribution to $f_F(E)=(e^{E/T_\nu}+1)^{-1}$ and take negligible hidden-fermion occupation. With these distributions held fixed, channel $i$ contributes $A_i(q)(1+f_S)$ for a massless scalar, where
\begin{equation}
 A_i(q)=\frac{m_i\Gamma_i^{(0)}T_\nu}{q^2}
 \ln\left[1+\exp\left(-\frac{q+m_i^2/(4q)}{T_\nu}\right)\right].
 \label{eq:gainzero}
\end{equation}
In the soft limit, set $u=m_i^2/(4qT_\nu)$. Maximizing $u^2\ln(1+e^{-u})$ gives
$2\ln(1+e^{-u_*})=u_* /(1+e^{u_*})$, with $u_*=2.12$.
Consequently the gain peaks at $q_*=m_i^2/(8.46T_\nu)$, with
$A_{i,\max}=8.15\Gamma_i^{(0)}(T_\nu/m_i)^3$.
At $T_\nu=1\,\mathrm{keV}$ the two peaks have momenta
$8.85\times10^{-9}\,\eV$ and $2.97\times10^{-7}\,\eV$, and gain coefficients $7.76\times10^3\,\mathrm{s}^{-1}$ and $231\,\mathrm{s}^{-1}$.
These coefficients describe growth in the selected momentum modes. Integrated neutrino conversion also depends on their phase-space measure and the evolving populations of the decay products.

When the occupation numbers of the decay products are negligible, the integrated spontaneous source of scalar variance follows by weighting \eqr{eq:collision} with inverse energy. Interchanging the parent-energy and scalar-momentum integrals gives
\begin{equation}
 \begin{aligned}
 Q_{I,i}^{\nu,\rm sp}
 &=\frac{m_i\Gamma_i^{(0)}}{2\pi^2}
   \int_{m_i}^\infty\dd E\,f_F(E)
    \ln\frac{E+\sqrt{E^2-m_i^2}}{E-\sqrt{E^2-m_i^2}}\\
 &\simeq\frac{m_i\Gamma_i^{(0)}T_\nu\ln2}{\pi^2}
          \ln\frac{\sqrt2T_\nu}{m_i},
 \qquad T_\nu\gg m_i .
 \end{aligned}
 \label{eq:spontaneousvariance}
\end{equation}
The logarithm on the upper line is the integral of $\dd q/q$ over the allowed momenta of the decay products at fixed parent energy. It retains the soft weighting needed for variance production, whereas the total conversion rate weights scalar number. At $T_\nu=1\,\mathrm{keV}$, the contributions from $\nu_2$ and $\nu_3$ are $4.50\times10^{-12}$ and $1.29\times10^{-10}\,\eV^2\,\mathrm{s}^{-1}$. These quantities measure variance production integrated over phase space. As the decay products accumulate, the full factor is $\mathcal F_i=f_{\nu_i}(1-f_{n_i})+f_S(f_{\nu_i}-f_{n_i})$, which includes inverse reactions and Pauli blocking.

A finite scalar mass changes the parent energies that can populate a given soft mode even while the two-body channel remains open. We use the finite-mass decay kinematics of ref.~\cite{Kaiser:1993lasing} and evaluate its collision integral for the fixed Fermi--Dirac parents. For a prescribed pole mass $0<\mu_S<m_i$, write $\omega_q=\sqrt{q^2+\mu_S^2}$ and $\beta_i=1-\mu_S^2/m_i^2$. At fixed $q$, the permitted parent energies lie in the finite interval
\begin{equation}
 E_\pm(q)=\frac{(m_i^2+\mu_S^2)\omega_q
                 \pm(m_i^2-\mu_S^2)q}{2\mu_S^2}.
 \label{eq:endpoints}
\end{equation}
The lower endpoint can also be written as
$E_-=q+(m_i^2+\mu_S^2)/[2(\omega_q+q)]$, which approaches
$q+m_i^2/(4q)$ as $\mu_S\to0$. The scalar collision term is
$m_i\Gamma_i^{(0)}\beta_i(\omega_q q)^{-1}
\int_{E_-}^{E_+}\dd E\,{\cal F}_i^{(\mu)}(E,E-\omega_q)$. In $\mathcal F_i^{(\mu)}$, the scalar occupation is evaluated at momentum $q=\sqrt{(E-k)^2-\mu_S^2}$; the fermion factors retain their arguments in \eqr{eq:statistical}.
For a fixed distribution of parent neutrinos, this gives
\begin{equation}
 \begin{aligned}
 A_i^{(\mu)}(q)&=
 \frac{m_i\Gamma_i^{(0)}\beta_iT_\nu}{\omega_q q}
 \ln\frac{1+e^{-E_-/T_\nu}}{1+e^{-E_+/T_\nu}},\\
 A_i^{(\mu)}(0)&=
 \Gamma_i^{(0)}\left(\frac{m_i}{\mu_S}\right)^3\beta_i^2
 f_F\left(\frac{m_i^2+\mu_S^2}{2\mu_S}\right),
 \end{aligned}
 \label{eq:gainmass}
\end{equation}
A massless scalar with $q\to0$ requires a parent energy tending to infinity. For finite $\mu_S$, zero-momentum emission instead samples the finite energy $(m_i^2+\mu_S^2)/(2\mu_S)$. This change can raise the soft gain before the shrinking phase space suppresses it. Integrating the corresponding parent loss term recovers the squared threshold factor in Eq.~\eqref{eq:scalarwidth}.

\begin{figure}[htbp]
 \centering
 \includegraphics[width=0.78\linewidth]{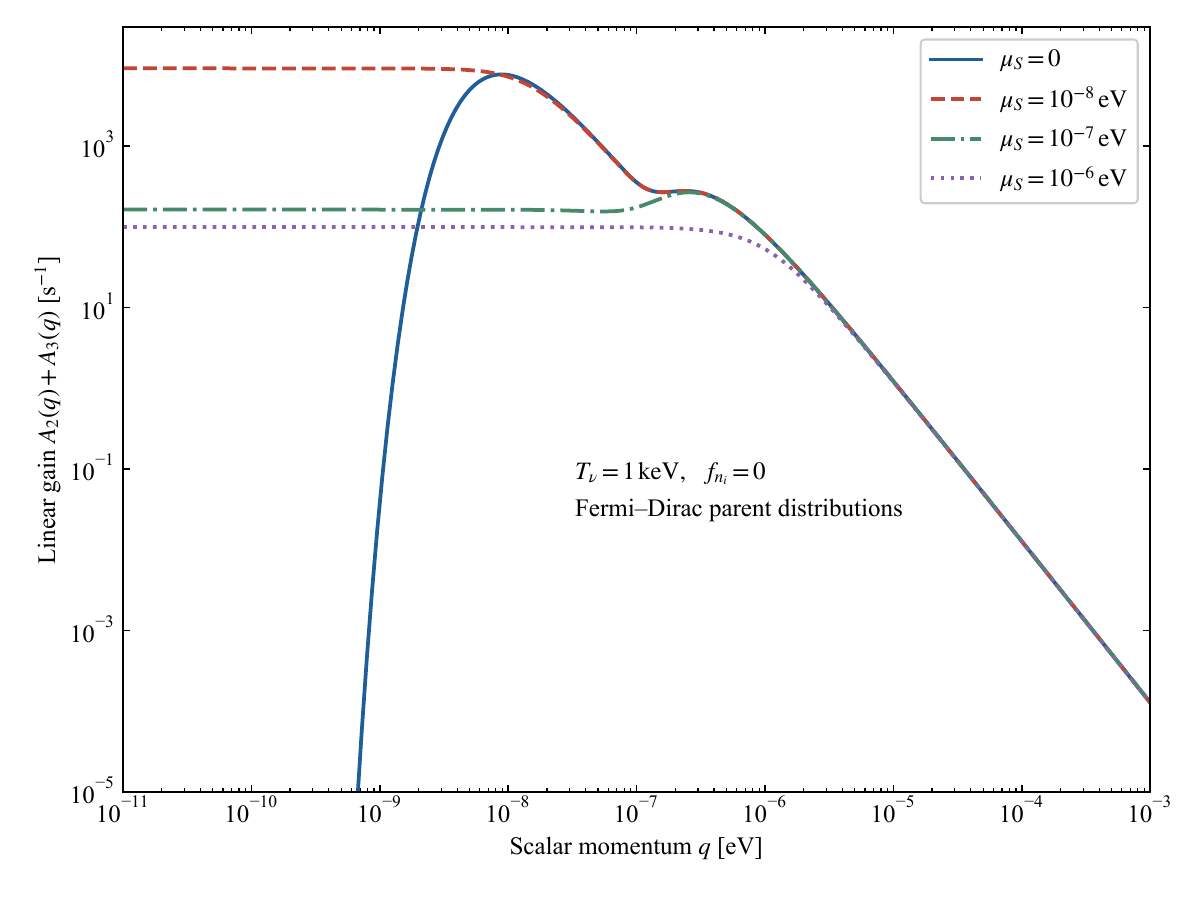}
 \caption{Linear coefficient $\sum_iA_i^{(\mu)}(q)$ multiplying $1+f_S$, evaluated for fixed thermal active-neutrino parents at $T_\nu=1\,\mathrm{keV}$ and empty hidden-fermion states. The four curves use prescribed constant scalar pole masses and the central vacuum couplings. The finite-mass curves include the parent-energy interval in \eqr{eq:endpoints}.}
 \label{fig:gain}
\end{figure}

Figure~\ref{fig:gain} shows the summed gain for four prescribed scalar masses. For $\mu_S=10^{-8}\,\eV$, the soft limit is $9.24\times10^3\,\mathrm{s}^{-1}$, exceeding the massless peak. For $\mu_S=10^{-7}\,\eV$, the maximum shifts to $q\simeq2.56\times10^{-7}\,\eV$ and becomes $269\,\mathrm{s}^{-1}$. At $\mu_S=10^{-6}\,\eV$, the soft limit is $100\,\mathrm{s}^{-1}$. Scalar dispersion therefore reshapes the gain nonmonotonically even when the vacuum two-body channels remain open.

\FloatBarrier
\subsection{Density dependence and scalar energy conservation}
We next ask how the production variance can generate a scalar mass in the range probed by these gain curves. For a charge-symmetric Gaussian state, the dispersion receives a quartic Hartree term and a retarded Yukawa self-energy. We distinguish the vacuum mass $m_{S,0}$, the general pole-mass variable $\mu_S^2\equiv\omega^2-q^2$, and the quartic-only self-consistent mass $m_{S,H}$ introduced below. On the scalar pole,
\begin{equation}
 \mu_S^2=m_{S,0}^2+4\lambda_S I_S+\Ren\Pi^R_{S,Y}(\omega,q),
 \label{eq:hartree}
\end{equation}
where $\Pi^R_{S,Y}$ is the retarded Yukawa self-energy; in general, the pole mass depends on momentum. The scalar term follows from contraction of the quartic interaction and gives $\lambda_ST_S^2/3$ for a massless thermal complex scalar. For the nonthermal distributions in Table~\ref{tab:moments}, it is determined by the sum $I_S^{\rm ref}=I_S^{\rm prim}+I_S^{LH}$. The resulting quartic mass contribution is
$\delta m_{S,\lambda}^2=4.96\times10^{-7}\lambda_ST_\nu^2$, a factor of $12.1$ above the primary contribution alone.

At $T_\nu=1\,\mathrm{keV}$, the comparison $\delta m_{S,\lambda}=q_{*,i}$ gives
$\lambda_S=1.58\times10^{-16}$ for $\nu_2$ and $1.78\times10^{-13}$ for $\nu_3$.

Table~\ref{tab:quarticgain} quantifies the response of the same fixed-parent
kernel to the primary and combined reference variances. The bath population
raises the quartic mass estimate by a factor of $3.47$. At the smaller
quartic it slightly increases the maximum gain; at the larger quartic it
reduces the maximum and shifts it away from zero momentum.
\begin{table}[htbp]
 \centering
 \begin{tabular}{lllll}
 \toprule
 $\lambda_S$ & Scalar reference & $\mu_S\ [\eV]$
 & $q_{\max}\ [\eV]$ & $\max_q\sum_iA_i^{(\mu)}\ [\mathrm{s}^{-1}]$\\
 \midrule
 $10^{-16}$ & Primary & $2.03\times10^{-9}$ & $8.73\times10^{-9}$ & $7.76\times10^3$\\
 $10^{-16}$ & Primary and bath & $7.04\times10^{-9}$ & $6.88\times10^{-9}$ & $7.98\times10^3$\\
 $10^{-13}$ & Primary & $6.41\times10^{-8}$ & $0$ & $543$\\
 $10^{-13}$ & Primary and bath & $2.23\times10^{-7}$ & $2.35\times10^{-7}$ & $247$\\
 \bottomrule
 \end{tabular}
 \caption{Quartic mass estimates from the production variance and fixed-parent gain maxima at
 $T_\nu=1\,\mathrm{keV}$. The masses use
 $\mu_S^2=4\lambda_S I_S^{\rm ref}$ with zero vacuum mass and Yukawa
 self-energy. The parent neutrinos retain their Fermi--Dirac distributions
 and the hidden-fermion states are empty. The reference variances use massless scalar dispersion.}
 \label{tab:quarticgain}
\end{table}

The induced mass also changes the inverse-energy weight. To isolate the scalar self-interaction, set $\Pi^R_{S,Y}=0$ and define the Hartree mass at fixed occupations by $m_{S,H}^2=m_{S,0}^2+4\lambda_S I_S(m_{S,H})$, with $\omega_p^2=p^2+m_{S,H}^2$, $m_{S,0}^2\ge0$ and $\lambda_S\ge0$.
Define $J_S=\int\dd^3p\,[f_S+f_{S^\dagger}]/[(2\pi)^3\,4\omega_p^3]$, so that
$\partial I_S/\partial m_{S,H}^2=-J_S$.
The derivative of the difference between the two sides of the gap equation with respect to $m_{S,H}^2$ is $1+4\lambda_SJ_S>0$. For finite integrals the solution is unique and lies between $m_{S,0}^2$ and $m_{S,0}^2+4\lambda_SI_S(m_{S,0})$. Its response to a change of occupations is
$\delta m_{S,H}^2=4\lambda_S\delta I_S|_{m_{S,H}}/(1+4\lambda_SJ_S)$.
This denominator reduces the mass response when newly produced modes have momenta comparable to the induced mass. To determine its behaviour as the quartic is reduced, consider an infrared occupation
$f_+(p)\equiv f_S(p)+f_{S^\dagger}(p)=f_{+,0}+O[(p/\Lambda)^\beta]$ with $\beta>0$. Writing $\mu=m_{S,H}$ gives the small-mass response
\begin{equation}
 \begin{aligned}
 J_S(\mu)&=\frac{f_{+,0}}{8\pi^2}\ln\frac{\Lambda}{\mu}+O(1),\\
 I_S(\mu)&=I_S(0)-\frac{f_{+,0}\mu^2}{8\pi^2}
                    \ln\frac{\Lambda}{\mu}+O(\mu^2),
 \end{aligned}
 \label{eq:hartreelog}
\end{equation}
where $\Lambda$ is a fixed momentum scale within the regular infrared region.
At fixed occupations with $I_S(0)>0$, $m_{S,0}=0$ and $\lambda_S>0$, the positive gap solution consequently obeys
\[
 m_{S,H}^2=4\lambda_S I_S(0)
 \left[1+O\!\left(\lambda_S[1+f_{+,0}\ln(1/\lambda_S)]\right)\right].
\]
Thus
$4\lambda_SJ_S(m_{S,H})\to0$ as $\lambda_S\to0^+$, although
$J_S(0)$ diverges when $f_{+,0}>0$. The response is evaluated at the positive gap root. This
fixed-occupation limit requires reassessment if amplification changes the
infrared distribution or produces a coherent component.

The variance evolution must also account for the changing inverse-energy weight. With
$C_+=(\partial_t-Hp\partial_p)(f_S+f_{S^\dagger})$ and
$Q_I^{(H)}=\int\dd^3p\,C_+/[(2\pi)^3\,2\omega_p]$, integration by parts gives
$\dot{I}_S+2HI_S=Q_I^{(H)}-J_S(\partial_t(m_{S,H}^2)+2Hm_{S,H}^2)$.
Using the gap equation yields
\begin{equation}
 (1+4\lambda_SJ_S)(\dot{I}_S+2HI_S)
    =Q_I^{(H)}-2Hm_{S,0}^2J_S .
 \label{eq:hartreemoment}
\end{equation}
For zero vacuum mass and collisionless evolution, $I_S\propto a^{-2}$ and $m_{S,H}\propto a^{-1}$ follow within this quasiparticle treatment. A nonzero vacuum mass introduces the transition toward the nonrelativistic dilution law. The distribution determines both $J_S$ and the mass-dependent collision moment.

To infer the scalar equation of state from this dispersion, the interaction energy must be included along with the quasiparticle energy. In the conserving Hartree treatment, the quartic contribution already counted in the effective mass requires a compensating interaction term~\cite{LopezNacir:2014hartree}. With the complex-scalar normalization used here,
$\rho_S=\rho_{S,\rm qp}-2\lambda_S I_S^2$ and
$P_S=P_{S,\rm qp}+2\lambda_S I_S^2$.
Differentiating the subtraction cancels the scalar Hartree work term
$\partial_t(m_{S,H}^2) I_S=4\lambda_S I_S\dot{I}_S$ in the quasiparticle energy equation.
At fixed renormalized couplings, the same subtraction gives $\rho_S-3P_S=2m_{S,H}^2I_S-8\lambda_SI_S^2=2m_{S,0}^2I_S$. Thus the zero-vacuum-mass theory has $\rho_S=3P_S$ within this Hartree closure: the quartic-induced quasiparticle mass changes decay kinematics while the physical scalar stress tensor retains its radiation equation of state.

The same conserving energy limits the mass that a fixed scalar population can support. For $m_{S,0}=0$ and $\lambda_S>0$, the pointwise inequality $\omega_p^2\ge m_{S,H}^2$ gives
\begin{equation}
 \rho_{S,\rm qp}\ge8\lambda_S I_S^2,\qquad
 \rho_S\ge6\lambda_S I_S^2,\qquad
 m_{S,H}^4\le\frac83\lambda_S\rho_S .
 \label{eq:hartreeenergy}
\end{equation}
Self-consistent dispersion therefore bounds the variance at fixed scalar energy. An inflaton-dependent mass or Yukawa self-energy contributes additional energy exchange, with the fermion dispersion evaluated in the same state.

The finite number of scalar quanta gives an independent limit on the Hartree mass, which can be combined with number conservation in the decay network. Since
$I_S\le n_S/(2m_{S,H})$, the gap equation gives
\begin{equation}
 m_{S,H}(m_{S,H}^2-m_{S,0}^2)\le2\lambda_Sn_S,
 \qquad
 m_{S,H}^3\le2\lambda_Sn_S\quad(m_{S,0}=0).
 \label{eq:hartreenumber}
\end{equation}
Within the decay network in Eq.~\eqref{eq:collision},
$a^3(n_S+n_{\nu_2}+n_{\nu_3})$ is conserved. Two initially thermal Majorana
species therefore give
$n_S/T_\nu^3\le3\zeta(3)/\pi^2$ up to the small initial scalar abundance,
where $T_\nu\propto a^{-1}$ is a reference scale after the distributions
become nonthermal. The resulting upper masses at $T_\nu=1\,\mathrm{keV}$ are
$4.18\times10^{-3}\,\eV$ for $\lambda_S=10^{-16}$ and
$4.18\times10^{-2}\,\eV$ for $\lambda_S=10^{-13}$.
Thus quartic feedback in this restricted network cannot reach either
neutrino threshold at the smaller coupling or the $\nu_3$ threshold at the
larger coupling. Changes to the soft gain can occur well below these
thresholds. Additional number-changing reactions or a coherent component
require separate accounting before applying the history-dependent number
bound.

\subsection{Validity and dispersion feedback}\label{sec:validity}
The relevant continuation question is whether dispersion changes before substantial neutrino conversion and whether a quasiparticle description survives that change. At $T_\nu=1\,\mathrm{keV}$, the Standard Model radiation reference with $g_*=3.36$ and
$T_\gamma=(11/4)^{1/3}T_\nu$ gives $H=7.43\times10^{-7}\,\mathrm{s}^{-1}$.
The massless peak coefficients obey $A_{2,\max}/H=1.04\times10^{10}$ and
$A_{3,\max}/H=3.11\times10^8$. Their growth-to-oscillation ratios are
$\hbar A_{2,\max}/q_{*,2}=5.77\times10^{-4}$ and
$\hbar A_{3,\max}/q_{*,3}=5.13\times10^{-7}$. These compare a local
occupation gain with expansion and oscillation; $A_i$ is not a computed
spectral width or angular relaxation rate.

Large occupation alone does not imply $\langle S\rangle\ne0$. A kinetic
continuation requires narrow spectral support, slowly varying dispersion
on the microscopic scale, controlled connected and anomalous correlations,
and collision rates evaluated on the evolving distributions. The negligible
scalar redistribution of the dilute production reference cannot be
extrapolated solely from small $\lambda_S$ after Bose amplification.
For a local estimate at the prescribed initial state, a massless shell of unit width in $\ln q$ near $q_{*,2}$ requires
$f_S\sim3\times10^{16}$ to add variance comparable to
$I_S^{\rm ref}=0.124\,\eV^2$. The frozen gain reaches that value in roughly
$5\,\mathrm{ms}$, while the associated scalar number is only of order
$10^{-17}$ of one thermal neutrino species. This local estimate terminates
when the extra variance changes the dispersion. It shows why negligible
parent depletion can coexist with important dispersion feedback.

A controlled continuation must also include the Yukawa retarded self-energy on the evolving state. Its dimensional scale is $g_\nu T_\nu=1.54\times10^{-9}\,\eV$ at $1\,\mathrm{keV}$, so it cannot be discarded uniformly as $\lambda_S\to0$. Figure~\ref{fig:gain} prescribes the scalar pole mass, while Table~\ref{tab:quarticgain} isolates the quartic contribution by setting this self-energy to zero. They establish sensitivity to the production variance, while its persistence through nonlinear conversion and its effect on angular transport remain open.

\section{Conclusions}\label{sec:conclusions}
For two separated heavy Majorana mass levels, the imaginary threshold overlap responsible for the inclusive mixed lepton-number-violating source also limits cancellation of the portal generated by their exchange. The resulting norm domain fixes the cancellation freedom at specified threshold strengths. With the stated ensemble and width normalization, it also gives a lower bound in terms of the normalized source and heavy masses. An independent matching contribution can remove this lower bound. With additional heavy levels, the relation survives at leading order in the heavy-mass hierarchy, with a remainder that depends on the individual thresholds.

The flavour realization isolates the bounded source while reproducing two nonzero neutrino masses. Its reflection family preserves the light mass matrix and redistributes the decay coupling between the massive modes. The maximal-source endpoint has equal-strength orthogonal coupling rows. On the specified reheating background, the baryon yield and primary hidden radiation fraction eliminate the portal normalizations and give conditional lower envelopes for weighted and individual vacuum widths.

The same operator produces scalars from the visible bath. At the benchmark, these scalars contribute $11.1$ times the primary scalar variance while carrying $4.81\times10^{-5}$ of its energy. Their softer spectrum explains this separation and makes the inverse-energy moment the relevant production quantity for scalar dispersion. This positive moment inherits the portal bound within the stated massless Born approximation and reheating history.

The production variance sets a scalar mass scale whose effect is tested with prescribed neutrino and hidden-fermion distributions. Conserving Hartree relations then constrain self-consistent scalar dispersion. Within the stated closure and reaction network, energy and number bound the induced scalar mass. The local feedback estimate identifies when dispersion can change before appreciable neutrino depletion. Scalar-potential parameters and the primordial state remain independent inputs, and radiative accuracy requires sequential matching and running. Determining the duration and saturation of the soft gain requires coupled evolution with inverse reactions, redistribution and retarded self-energies. Cosmological free streaming also requires angular perturbations; the benchmark's cosmological viability remains open.

\appendix
\section{Decay amplitudes, production and scalar matching}\label{app:cut}
\subsection{The heavy-particle cut}\label{app:heavycut}
The cut fixes the normalization and CP contractions of the mixed source used in Section~\ref{sec:bound}. The hidden part of the heavy two-point function follows from the coupling in \eqr{eq:lag}. In two-component notation its loop integral has the form
\begin{equation}
 i\Sigma^{nS}_{IJ}(p)=
 2D_{IJ}\int\frac{\dd^4k}{(2\pi)^4}
       \frac{ik\cdot\sigma}{k^2+i0}\,
       \frac{i}{(p-k)^2+i0}.
 \label{eq:selfenergy}
\end{equation}
The factor two comes from the $\sqrt2$ in Eq.~\eqref{eq:lag}. The physical massless cut has
$\int\dd\Phi_2=1/(8\pi)$ and
$\int\dd\Phi_2\,k^\mu=p^\mu/(16\pi)$.
With the Cutkosky discontinuity convention~\cite{Cutkosky:1960sp}, the absorptive self-energy carries the factor $D_{IJ}/(16\pi)$ multiplying $p\cdot\sigma$. Including the conjugate Majorana flow gives the pole width in \eqr{eq:widthheavy}. The off-diagonal interference with the visible tree amplitude produces $A[D]$ and $B[D]$ in \eqr{eq:kernels}, with the self-energy loop functions $\FS$ and $\GS$ in \eqr{eq:source}. The hidden fields enter this self-energy through their on-shell states; the visible vertex contribution retains its usual lepton--Higgs cut.

The inverse reactions and the subtraction of real intermediate heavy states organize the equilibrium cancellation~\cite{Garny:2010nz,Beneke:2010ctp,Garbrecht:2011ctp,Garbrecht:2018mrp}. In a scattering channel with an on-shell heavy contribution the subtracted reaction density is
$\gamma^{\rm sub}_{ab\to cd}=\gamma_{ab\to cd}-\gamma_{ab\to X_1}\operatorname{Br}(X_1\to cd)$.
Combining it with decays and inverse decays gives a vanishing asymmetry source in thermal equilibrium. This optical and kinetic normalization is the basis of \eqr{eq:source}.

\subsection{The bath-production spectrum}\label{app:bathspectrum}
The differential production kernel determines which scalar momenta carry the variance in Section~\ref{sec:bath}. Its angular weight matters because the variance emphasizes low outgoing energies. The single-component contact cross section is $|C|^2/(32\pi)$. The squared $\sqrt2$ portal factor, weak components and charge conjugation give the inclusive factor eight. The Weyl contraction has squared amplitude proportional to $p_L\cdot p_n$, giving the normalized angular distribution $(1+c_*)/2$ for the outgoing scalar. For incident energies $xT$ and $zT$ and incident angle cosine $c$, set $w=x+z$, $d=(x-z)/w$ and $\beta^2=1-2xz(1-c)/w^2$. After azimuthal integration the probability in the boost-direction cosine $u$ is $\dd P/\dd u=[1+(d/\beta)u]/2$, with $E_S/T=w(1+\beta u)/2$. Thus $\langle E_S/T\rangle=(2x+z)/3$ and $\langle E_n/T\rangle=(x+2z)/3$.

The inverse-energy average is
\begin{equation}
 \begin{aligned}
 \mathcal W(x,z,c)&\equiv T\left\langle\frac1{2E_S}\right\rangle
 =\frac1{2w}\left[L(\beta)+\frac d{\beta^2}\{2-L(\beta)\}\right],\\
 L(\beta)&=\frac1\beta\ln\frac{1+\beta}{1-\beta},\\
 c_I&=\frac1{16\pi^5}
 \int_0^\infty\frac{x^2\,\dd x}{e^x+1}
 \int_0^\infty\frac{z^2\,\dd z}{e^z-1}
 \int_{-1}^1\frac{\dd c}{2}(1-c)\mathcal W(x,z,c).
 \end{aligned}
 \label{eq:inverseproduction}
\end{equation}
The small-$\beta$ limit is continuous. The same incident measure with weights $1$, $T(2x+z)/3$ and $Tw$ gives $Q_n$, $Q_{\rho_S}$ and $Q_D$ in Section~\ref{sec:bath}, respectively. In particular, the integral multiplying $1/(16\pi^5)$ in \eqr{eq:inverseproduction} is $1.080$.

The differential spectrum follows by integrating over the two scattering angles. For $\ell=p_S/T$, set $r_\ell=2\ell/w-1$ and $b_*=\max(|d|,|r_\ell|)$. The kinematic support requires $w\ge\ell$, and the incident-angle integration covers $b_*\le\beta\le1$. It gives
\begin{equation}
 \begin{aligned}
 \mathcal S(\ell)
 &=\frac{\ell}{4}\int_0^\infty\dd x\int_0^\infty\dd z\,
 \frac{w^3(1-b_*)^2}{(e^x+1)(e^z-1)}
 \left[\frac{b_*+2}{3}+\frac{dr_\ell}{b_*}\right]\Theta(w-\ell),\\
 \left.\frac{\dd\dot n_S^{LH}}{\dd\ln p}\right|_{\rm production}
 &=\frac{\mathcal C^2T^6}{16\pi^5}\mathcal S(p/T).
 \end{aligned}
 \label{eq:productionshape}
\end{equation}
The ratio $dr_\ell/b_*$ is defined by continuity at $b_*=0$. The shape has the moment integrals
$\int\dd\ln\ell\,\mathcal S=3\zeta(3)^2$,
$\int\dd\ln\ell\,\ell\mathcal S=\pi^4\zeta(3)/9$ and
$\int\dd\ln\ell\,\mathcal S/\ell=32\pi^5c_I$.
These recover the independently weighted sources in \eqr{eq:bathmoments}. For Maxwell--Boltzmann incident distributions, the angular contribution odd under $x\leftrightarrow z$ integrates to zero. The remaining integral gives $\mathcal S_{\rm MB}(\ell)=\ell^3\int_\ell^\infty\dd w\,e^{-w}(w-\ell)^2=2\ell^3e^{-\ell}$.

For $\ell\to0$, the leading term is
$\mathcal S(\ell)=(\pi^2\ln2/3)\ell^3+o(\ell^3)$.
Consequently the charge-symmetric occupation per scalar state approaches
\begin{equation}
 f_S^{LH}(y\to0)=
 \frac{\mathcal C^2\ln2}{48\pi}
       \int\dd\ln a\,\frac{T^3}{H}=2.57\times10^{-5}.
 \label{eq:bathoccupation}
\end{equation}
The variance below $y_c$ consequently obeys $I_S^{LH}(0<y<y_c)/T_\nu^2=f_S^{LH}(0)y_c^2/(4\pi^2)+o(y_c^2)$. Modes with $y<10^{-3}$ carry approximately $5.7\times10^{-6}$ of the bath variance. The enhanced variance is therefore carried by the softer bulk population, whose occupation remains dilute in the massless production treatment.

\subsection{The perturbative production regime}\label{app:regime}
Independent redshifting during production requires scattering and inverse decays to remain rare. We bound their integrated probabilities using the total hidden target density, including scalars and fermions with both charges. The primary contribution is $n_D^{\rm prim}=4B_Xb_D\rho_0(1-\Phi)/(m_\phi a^3)$, and the bath contribution obeys $\dd(a^3n_D^{LH})/\dd\ln a=2a^3Q_n/H$. The integrated target density is $\int\dd t\,n_D=2.55\times10^{22}\,\GeV^2$.

For incident energies $p$ and $k$, isotropic narrow-width averaging gives
$\langle\sigma v\rangle_{nS\to X_1}=\pi\eta\kappa_1M_1^2\Theta(4pk-M_1^2)/(4p^2k^2)\le4\pi\eta\kappa_1/M_1^2$.
The Majorana spin sum and the $\sqrt2$ portal normalization are included.
The dimension-five $C_{nn}$ contact vertex is $4C_{nn}^{rs}$; its inclusive crossed reactions have the cross-section bound $\norm{C_{nn}}^2/(2\pi)$ after the identical-particle factors are included. The contact coefficient is dominated by the heavier threshold, while resonant production of $X_1$ is estimated separately. Counting every hidden particle as a target gives conservative upper bounds on the time-integrated reaction rates per incident particle in Table~\ref{tab:regime}. These small values support independent redshifting during production.

\begin{table}[htbp]
 \centering

 \begin{tabular}{ll}
 \toprule
 Quantity & Value or upper bound\\
 \midrule
 Inverse decay $nS\to X_1$ & $<7.2\times10^{-5}$\\
 Dimension-five $C_{nn}$ scattering & $<2.7\times10^{-5}$\\
 Weinberg-operator washout & $6.54\times10^{-6}$\\
 Upper bound on heavy-particle occupation & $1.75\times10^{-7}$\\
 Maximum primary scalar occupation & $1.85\times10^{-8}$\\
 Infrared bath-produced scalar occupation & $2.57\times10^{-5}$\\
 \bottomrule
 \end{tabular}
 \caption{Time-integrated reaction rates per incident particle and occupation numbers during the specified perturbative era. The hidden bounds count both production components and all hidden targets. Occupations are per populated spin or charge state; the scalar values follow from the production spectra.}
 \label{tab:regime}
\end{table}

For the scalar-curvature estimate we write $M_1(\phi)=M_1+\lambda_X\phi/\sqrt2$.
The partial width is $\Gamma_{\phi\to XX}=\lambda_X^2m_\phi\beta_X^3/(16\pi)$.
The branching fraction fixed by the baryon abundance gives $\lambda_X=7.40\times10^{-7}$.
The isolated lighter-block curvature before subtracting low powers of the
inflaton is
\begin{equation}
 \Delta m_S^2(\phi)=-\frac{\Tr(D_{11})}{4\pi^2}M_1(\phi)^2
       \left[\ln\frac{M_1(\phi)^2}{\mu_R^2}-1\right],
 \quad M_1(\phi)=M_1+\lambda_X\phi/\sqrt2.
 \label{eq:inflatonportal}
\end{equation}
At $\mu_R=M_1$, its quadratic coefficient is
$\Delta\kappa_{\phi S}^{\rm CW}
=-\Tr(D_{11})\lambda_X^2/(4\pi^2)=-9.68\times10^{-20}$.
The magnitude lies near the chosen benchmark scale, while the positive
renormalized value is a separate matching choice. The constant, linear and
quadratic curvature coefficients are fixed before estimating the residual
cubic term. The isolated linear term vanishes at this subtraction scale;
the corresponding operator is allowed by the stated symmetry.

After these terms through $\phi^2$ are absorbed into the matching coefficients, the leading remainder is
$-\Tr(D_{11})(\lambda_X\phi/\sqrt2)^3/(6\pi^2M_1)$.
Its maximum magnitude is $4.03\times10^{-8}\,\GeV^2$, compared with $0.447\,\GeV^2$ from the positive $\kappa_{\phi S}\phi^2S^\dagger S$ interaction. The ratio is $9.00\times10^{-8}$ at the initial amplitude and decreases as the condensate redshifts. The retained positive curvature therefore controls the specified perturbative background.

\subsection{Isolated heavier-block scalar thresholds}\label{app:quartic}
These threshold terms quantify the heavy-loop scales that must be balanced by the scalar matching inputs in Section~\ref{sec:running}. For the texture in \eqr{eq:texture}, write $d_2=c_U^2\kappa_2/\eta$ and
$G_2=\sqrt2\lambda_2=\sqrt{2d_2}F^*$. The unitary hidden rotation
$n=F^Tn'$ gives $G_2F^T=\sqrt{2d_2}\id_2$ and leaves the heavy Majorana
mass unchanged. For constant backgrounds $h^2=H^\dagger H$ and
$s^2=S^\dagger S$, the heavy-to-light mixing obeys
$BB^\dagger=(\kappa_2h^2+2d_2s^2)\id_2\equiv u\id_2$.
The two equal heavy singular values satisfy
$M_{2,+}^2=M_2^2+2u-u^2/M_2^2+O(u^3/M_2^4)$.

A fermionic eigenmode contributes
$-m^4[\ln(m^2/\mu_R^2)-3/2]/(32\pi^2)$ to the
$\overline{\mathrm{MS}}$ effective potential~\cite{Coleman:1973jx}.
Writing $z=u/M_2^2$ and $L=\ln(M_2^2/\mu_R^2)$, the coefficient of
$z^2$ in $M_{2,+}^4[\ln(M_{2,+}^2/\mu_R^2)-3/2]/M_2^4$ is $2L+2$.
The light eigenvalues first contribute at eighth order in the backgrounds.
Summing the two heavy modes therefore gives
\begin{equation}
 \Delta V_{X_2}^{(4)}=-\frac{1+\ln(M_2^2/\mu_R^2)}{8\pi^2}
       (\kappa_2H^\dagger H+2d_2S^\dagger S)^2 .
 \label{eq:heavypotential}
\end{equation}
The additive scalar and mixed quartic coefficients are
\begin{align}
 \Delta\lambda_{S,\rm CW}^{(X_2)}(\mu_R)
 &=-\frac{\Tr(D_{22}^2)}{4\pi^2}
    \left[1+\ln\frac{M_2^2}{\mu_R^2}\right],
 \label{eq:quarticthreshold}\\
 \Delta\lambda_{HS}^{(X_2),\rm CW}(\mu_R)
 &=-\frac{\Tr(K_{22}D_{22})}{4\pi^2}
       \left[1+\ln\frac{M_2^2}{\mu_R^2}\right].
 \label{eq:higgsportal}
\end{align}
At $\mu_R=M_2$, these are $-2.57\times10^{-2}$ and
$-3.10\times10^{-4}$, respectively. These isolated potential coefficients
specify the heavy-loop scales against which the chosen renormalized
couplings are compared. Canonical effective-theory matching also includes
field normalization, the lighter block and independent threshold terms.

\section{Flavour information in the reflection family}\label{app:reflection}
The source also has a flavour component whose size is not fixed by its trace.
Writing $s_\theta=\sin\theta$ and $c_\theta=\cos\theta$, direct evaluation
of Eq.~\eqref{eq:kernels} gives
\begin{equation}
 \eps=\frac{c_U\kappa_2}{16\pi(1+\eta)}W^\dagger
 \left[\FS\xi\id_2+\GS P_\theta\right]W,
 \qquad P_\theta=
 \begin{pmatrix}
 -\xi&(1-i)s_\theta c_\theta\\
 (1+i)s_\theta c_\theta&\xi
 \end{pmatrix}.
 \label{eq:reflectionfullsource}
\end{equation}
The traceless contribution has eigenvalues proportional to
$\pm\sqrt{\xi(2-\xi)}$, while the inclusive source is proportional to
$\xi$. The trace determines the baryon normalization in the weak-washout
regime considered here. A flavour-dependent washout extension must retain
this matrix information.

The two squared singular values of $G_{\rm m}$ follow from Eq.~\eqref{eq:reflectioncouplings} and are $2v^2[x^2+y^2\pm2xy\sqrt{1-\xi^2}]$, with
\begin{equation}
 \det(G_{\rm m}G_{\rm m}^\dagger)=(2v^2)^2[(x^2-y^2)^2+4x^2y^2\xi^2].
 \label{eq:reflectiondet}
\end{equation}
Thus a nonzero mixed source also keeps both singular couplings nonzero in this family.

\section{Orientation deformations of the aligned texture}\label{app:orientation}
This extension quantifies how a conventional visible source returns when the internal orientation leaves the source-isolating reflection family. The heavy widths remain isotropic, so this deformation separates source restoration from a possible splitting of the degenerate pairs.
Within the family $F=e^{i\pi/4}\boldsymbol n\cdot\boldsymbol\sigma$, with
$\boldsymbol n$ real and of unit norm, define
$\delta=1-2n_y^2$ and
$\mathsf H_\delta=2n_y(n_z\sigma_1-n_x\sigma_3)$. Then
$FF^T=-\mathsf H_\delta+i\delta\id_2$, and the source is
\begin{equation}
 \eps=\frac{\kappa_2}{16\pi(1+\eta)}W^\dagger
 \left\{\bigl[(\FV+\FS)\delta+c_U\FS\bigr]\id_2
       +c_U\GS\mathsf H_\delta\right\}W .
 \label{eq:orientationsource}
\end{equation}
The ratio of the visible and mixed inclusive contributions is
$\delta(\FV+\FS)/(c_U\FS)$, approaching $3\delta/(2c_U)$ in the heavy
hierarchy. The squared light masses satisfy
\[
 m_{3,2}^2=v^4[h_1^2+h_2^2\pm2h_1h_2\sqrt{1-\delta^2}].
\]
Keeping the measured masses while varying this orientation therefore requires
\begin{equation}
 \begin{gathered}
 |\delta|\leq\frac{2m_2m_3}{m_3^2+m_2^2}=0.335,\\
 h_1^2,h_2^2=\frac12\left[\Sigma\pm
 \sqrt{\Sigma^2-\frac{\Delta^2}{1-\delta^2}}\right],\qquad
 \Sigma=\frac{m_3^2+m_2^2}{2v^4},\qquad
 \Delta=\frac{m_3^2-m_2^2}{2v^4}.
 \end{gathered}
 \label{eq:orientationdomain}
\end{equation}
Here $h_I=\kappa_I/M_I$ as in Section~\ref{sec:classification}. The branch $h_1\ge h_2$ connects continuously to the central texture.

\section{A discrete realization of the portal phase}\label{app:phase}
The endpoint texture uses a specific relative portal phase. The construction below shows how a discrete symmetry and a chosen vacuum can select that phase, while leaving the Yukawa row normalizations and the scalar hierarchy as matching inputs. Let $\omega=e^{i\pi/4}$, $D_8=\diag(\omega,\omega^5)$ and
$V=\left(\begin{smallmatrix}c&is\\is&c\end{smallmatrix}\right)$, with
$c=\cos(\pi/8)$ and $s=\sin(\pi/8)$. Then $F=VD_8V^\dagger$.
The generator in the heavy-field mass basis is $G_8=V^*D_8V^T$.
Writing $\widehat X_2=V^TX_2$ for the fields with definite $Z_8$ charges, the mass transforms by congruence:
\begin{equation}
 \widehat M_2=V^\dagger(M_2\id_2)V^*
 =\frac{M_2}{\sqrt2}\begin{pmatrix}1&-i\\-i&1\end{pmatrix},
 \qquad V\widehat M_2V^T=M_2\id_2.
 \label{eq:congruence}
\end{equation}
The two Takagi singular values are therefore $M_2$.

Assign charges $(1,5,7,3,6,2)$ modulo eight to
$(\widehat X_{2,+},\widehat X_{2,-},\zeta_7,\zeta_3,\Phi_6,\Phi_2)$.
The mass interaction is
$-[y_6\Phi_6(\widehat X_{2,+}^2+\widehat X_{2,-}^2)
+2y_2\Phi_2\widehat X_{2,+}\widehat X_{2,-}]/(2\sqrt2)+\mathrm{h.c.}$
The matching choice $y_6=y_2$, $y_6f_M=M_2$, and the vacuum
$\Phi_6=f_M$, $\Phi_2=-if_M$ give \eqr{eq:congruence}.
Visible rows arise from $\zeta_7\widehat X_{2,+}(L\cdot H)$ and
$\zeta_3\widehat X_{2,-}(L\cdot H)$.

Define $\Phi_M=2^{-1/2}\left(\begin{smallmatrix}\Phi_6&\Phi_2\\\Phi_2&\Phi_6\end{smallmatrix}\right)$.
The radial terms fix the field magnitudes, the $\lambda_{\rm iso}$ term keeps the two heavy singular values equal, and the remaining terms select discrete phases. For positive coefficients, one CP-invariant potential is
\begin{equation}
 \begin{aligned}
 V_{\rm phase}={}&
 \lambda_r\sum_{q=2,6}(|\Phi_q|^2-f_M^2)^2
 +\frac{\lambda_4}{f_M^4}\sum_{q=2,6}|\Phi_q^4-f_M^4|^2
 +\frac{\lambda_{\rm rel}}{f_M^4}|(\Phi_2\Phi_6^*)^2+f_M^4|^2\\
 &+\lambda_{\rm iso}\Tr[(\Phi_M^\dagger\Phi_M-f_M^2\id_2)^2]
 +\lambda_\zeta\sum_{q=3,7}(|\zeta_q|^2-f_\zeta^2)^2
 +\frac{\lambda_8}{f_\zeta^{12}}\sum_{q=3,7}|\zeta_q^8-f_\zeta^8|^2\\
 &+\frac{\lambda_{37}}{f_\zeta^4}|(\zeta_7\zeta_3^*)^2-f_\zeta^4|^2
 +\lambda_\times\left|\zeta_7\zeta_3^*
                  +\frac{f_\zeta^2}{f_M^2}\Phi_6^2\right|^2
 +\lambda_c\sum_{q=3,7}
          \left|\zeta_q^2+\frac{f_\zeta^2}{f_M}\Phi_2^*\right|^2 .
 \end{aligned}
 \label{eq:potential}
\end{equation}
Every term is invariant under $Z_8$. At the chosen vacuum values of $\Phi_2$ and $\Phi_6$ the cross term fixes
$\zeta_3=-\zeta_7$. For $\lambda_c=0$, eight portal branches
$\zeta_7=f_\zeta\omega^k$ remain. Imposing the alignment in Eq.~\eqref{eq:texture} on each branch gives both $Y_2$ and $\lambda_2$ the phase $e^{i(k-7)\pi/4}$, so the relative mixed source is $\cos[(k-7)\pi/2]$. The final term imposes $\zeta_q^2=-if_\zeta^2$ and selects
$k=3,7$ for these vacuum values. Including the mass phases gives sixteen minima in two CP-related $Z_8$ orbits. Conjugation of the full portal coefficients reverses the source sign.

Neutral messenger pairs can generate the visible rows from
$\Lambda_rP_r\overline P_r+g_r\zeta_r\widehat X_{2,r}P_r
+h_{r\alpha}\overline P_r(L_\alpha\cdot H)$.
Choosing $g_rh_{r\alpha}/\Lambda_r$ proportional to the rows of $V^\dagger W$
and inserting the selected phases yields
$\widehat Y_2=\sqrt{\kappa_2}D_8^\dagger V^\dagger W$ and hence \eqr{eq:texture}.
The messenger rows, common mass coefficient and hidden normalization remain matching inputs. The allowed operators $S^\dagger S$, $(S^\dagger S)^2$ and $H^\dagger HS^\dagger S$ leave the scalar hierarchy and negligible Higgs portal unprotected. We assume that symmetry breaking precedes inflation, which selects a homogeneous vacuum, and that radial and messenger thresholds lie above the maximum reheating temperature.

\section*{Conflict of interest}
The author declares no conflicts of interest.

\section*{Data Availability}
This study is theoretical. No data were generated or analyzed.

\bibliographystyle{JHEP}
\bibliography{references}

@article{Antusch:2005gp,
    title = {{Running Neutrino Mass Parameters in See-Saw Scenarios}},
    volume = {03},
    ISSN = {1029-8479},
    url = {http://dx.doi.org/10.1088/1126-6708/2005/03/024},
    doi = {10.1088/1126-6708/2005/03/024},
    number = {03},
    journal = {JHEP},
    publisher = {Springer Science and Business Media LLC},
    author = {Antusch, Stefan and Kersten, J{\"o}rn and Lindner, Manfred and Ratz, Michael and Schmidt, Michael Andreas},
    year = {2005},
    month = {Mar},
    pages = {024},
    eprint = {hep-ph/0501272},
    archivePrefix = {arXiv},
    primaryClass = {hep-ph}
}

@article{Blanchet:2011density,
    author = {Blanchet, Steve and Di Bari, Pasquale and Jones, David A. and Marzola, Luca},
    title = {{Leptogenesis with heavy neutrino flavours: from density matrix to Boltzmann equations}},
    eprint = {1112.4528},
    archivePrefix = {arXiv},
    primaryClass = {hep-ph},
    doi = {10.1088/1475-7516/2013/01/041},
    journal = {JCAP},
    volume = {01},
    pages = {041},
    year = {2013}
}

@article{Datta:2022reheating,
    author = {Datta, Arghyajit and Roshan, Rishav and Sil, Arunansu},
    title = {{Effects of Reheating on Charged Lepton Yukawa Equilibration and Leptogenesis}},
    eprint = {2206.10650},
    archivePrefix = {arXiv},
    primaryClass = {hep-ph},
    doi = {10.1103/physrevlett.132.061802},
    journal = {Phys. Rev. Lett.},
    volume = {132},
    number = {6},
    pages = {061802},
    year = {2024}
}

@article{Cutkosky:1960sp,
    author = {Cutkosky, R. E.},
    title = {{Singularities and discontinuities of Feynman amplitudes}},
    doi = {10.1063/1.1703676},
    journal = {J. Math. Phys.},
    volume = {1},
    pages = {429--433},
    year = {1960}
}

@article{Sakharov:1967dj,
    author = {Sakharov, A. D.},
    title = {{Violation of CP Invariance, C asymmetry, and baryon asymmetry of the universe}},
    doi = {10.1070/pu1991v034n05abeh002497},
    journal = {Pisma Zh. Eksp. Teor. Fiz.},
    volume = {5},
    pages = {32--35},
    year = {1967},
    note = {Reprinted in Sov. Phys. Usp. 34 (1991) 392--393}
}

@article{tHooft:1976rip,
    author = {'t Hooft, Gerard},
    editor = {Shifman, Mikhail A.},
    title = {{Symmetry Breaking Through Bell-Jackiw Anomalies}},
    reportNumber = {PRINT-76-0254 (HARVARD)},
    doi = {10.1103/physrevlett.37.8},
    journal = {Phys. Rev. Lett.},
    volume = {37},
    pages = {8--11},
    year = {1976}
}

@article{Kuzmin:1985mm,
    author = {Kuzmin, V. A. and Rubakov, V. A. and Shaposhnikov, M. E.},
    title = {{On the Anomalous Electroweak Baryon Number Nonconservation in the Early Universe}},
    reportNumber = {IC/85/8},
    doi = {10.1016/0370-2693(85)91028-7},
    journal = {Phys. Lett. B},
    volume = {155},
    pages = {36--42},
    year = {1985}
}

@article{Manton:1983nd,
    author = {Manton, N. S.},
    title = {{Topology in the Weinberg-Salam Theory}},
    reportNumber = {NSF-ITP-83-64},
    doi = {10.1103/physrevd.28.2019},
    journal = {Phys. Rev. D},
    volume = {28},
    pages = {2019--2026},
    year = {1983}
}

@article{Klinkhamer:1984di,
    author = {Klinkhamer, Frans R. and Manton, N. S.},
    title = {{A Saddle Point Solution in the Weinberg-Salam Theory}},
    reportNumber = {NSF-ITP-84-57},
    doi = {10.1103/physrevd.30.2212},
    journal = {Phys. Rev. D},
    volume = {30},
    pages = {2212--2220},
    year = {1984}
}

@article{Harvey:1990qw,
    author = {Harvey, Jeffrey A. and Turner, Michael S.},
    title = {{Cosmological Baryon and Lepton Number in the Presence of Electroweak Fermion Number Violation}},
    reportNumber = {FERMILAB-PUB-90-049-A, EFI-90-33},
    doi = {10.1103/physrevd.42.3344},
    journal = {Phys. Rev. D},
    volume = {42},
    pages = {3344--3349},
    year = {1990}
}

@article{Weinberg:1979bt,
    author = {Weinberg, Steven},
    title = {{Cosmological Production of Baryons}},
    reportNumber = {HUTP-78/A040},
    doi = {10.1103/physrevlett.42.850},
    journal = {Phys. Rev. Lett.},
    volume = {42},
    pages = {850--853},
    year = {1979}
}

@article{Weinberg:1979operator,
    author = {Weinberg, Steven},
    title = {{Baryon- and Lepton-Nonconserving Processes}},
    doi = {10.1103/physrevlett.43.1566},
    journal = {Phys. Rev. Lett.},
    volume = {43},
    pages = {1566--1570},
    year = {1979}
}

@article{Nanopoulos:1979gx,
    author = {Nanopoulos, Dimitri V. and Weinberg, Steven},
    title = {{Mechanisms for Cosmological Baryon Production}},
    reportNumber = {HUTP-79/A023},
    doi = {10.1103/physrevd.20.2484},
    journal = {Phys. Rev. D},
    volume = {20},
    pages = {2484--2493},
    year = {1979}
}

@article{Fukugita:1986hr,
    author = {Fukugita, M. and Yanagida, T.},
    title = {{Baryogenesis Without Grand Unification}},
    reportNumber = {RIFP-641},
    doi = {10.1016/0370-2693(86)91126-3},
    journal = {Phys. Lett. B},
    volume = {174},
    pages = {45--47},
    year = {1986}
}

@article{Luty:1992un,
    author = {Luty, M. A.},
    title = {{Baryogenesis via leptogenesis}},
    doi = {10.1103/physrevd.45.455},
    journal = {Phys. Rev. D},
    volume = {45},
    pages = {455--465},
    year = {1992}
}

@article{Flanz:1994yx,
    author = {Flanz, Marion and Paschos, Emmanuel A. and Sarkar, Utpal},
    title = {{Baryogenesis from a lepton asymmetric universe}},
    eprint = {hep-ph/9411366},
    archivePrefix = {arXiv},
    reportNumber = {DO-TH-94-15},
    doi = {10.1016/0370-2693(94)01555-q},
    journal = {Phys. Lett. B},
    volume = {345},
    pages = {248--252},
    year = {1995},
    note = {[Erratum: Phys.Lett.B 384, 487--487 (1996), Erratum: Phys.Lett.B 382, 447--447 (1996)]}
}

@article{Covi:1996wh,
    author = {Covi, Laura and Roulet, Esteban and Vissani, Francesco},
    title = {{CP violating decays in leptogenesis scenarios}},
    eprint = {hep-ph/9605319},
    archivePrefix = {arXiv},
    reportNumber = {SISSA-66-96-EP, IC-96-73},
    doi = {10.1016/0370-2693(96)00817-9},
    journal = {Phys. Lett. B},
    volume = {384},
    pages = {169--174},
    year = {1996}
}

@article{Pilaftsis:1997jf,
    author = {Pilaftsis, Apostolos},
    title = {{CP violation and baryogenesis due to heavy Majorana neutrinos}},
    eprint = {hep-ph/9707235},
    archivePrefix = {arXiv},
    reportNumber = {MPI-PHT-97-30},
    doi = {10.1103/physrevd.56.5431},
    journal = {Phys. Rev. D},
    volume = {56},
    pages = {5431--5451},
    year = {1997}
}

@article{Pilaftsis:2003gt,
    author = {Pilaftsis, Apostolos and Underwood, Thomas E. J.},
    title = {{Resonant leptogenesis}},
    eprint = {hep-ph/0309342},
    archivePrefix = {arXiv},
    reportNumber = {MC-TH-2003-09},
    doi = {10.1016/j.nuclphysb.2004.05.029},
    journal = {Nucl. Phys. B},
    volume = {692},
    pages = {303--345},
    year = {2004}
}

@article{Davidson:2002qv,
    author = {Davidson, Sacha and Ibarra, Alejandro},
    title = {{A Lower bound on the right-handed neutrino mass from leptogenesis}},
    eprint = {hep-ph/0202239},
    archivePrefix = {arXiv},
    reportNumber = {OUTP-02-10P, IPPP-02-16, DCPT-02-32},
    doi = {10.1016/s0370-2693(02)01735-5},
    journal = {Phys. Lett. B},
    volume = {535},
    pages = {25--32},
    year = {2002}
}

@article{Buchmuller:2004nz,
    author = {Buchmuller, W. and Di Bari, P. and Plumacher, M.},
    title = {{Leptogenesis for pedestrians}},
    eprint = {hep-ph/0401240},
    archivePrefix = {arXiv},
    reportNumber = {DESY-03-100, UAB-FT-551, CERN-TH-2003-199},
    doi = {10.1016/j.aop.2004.02.003},
    journal = {Annals Phys.},
    volume = {315},
    pages = {305--351},
    year = {2005}
}

@article{Buchmuller:2005eh,
    author = {Buchmuller, W. and Peccei, R. D. and Yanagida, T.},
    title = {{Leptogenesis as the origin of matter}},
    eprint = {hep-ph/0502169},
    archivePrefix = {arXiv},
    reportNumber = {DESY-05-031},
    doi = {10.1146/annurev.nucl.55.090704.151558},
    journal = {Ann. Rev. Nucl. Part. Sci.},
    volume = {55},
    pages = {311--355},
    year = {2005}
}

@article{Davidson:2008bu,
    author = {Davidson, Sacha and Nardi, Enrico and Nir, Yosef},
    title = {{Leptogenesis}},
    eprint = {0802.2962},
    archivePrefix = {arXiv},
    primaryClass = {hep-ph},
    doi = {10.1016/j.physrep.2008.06.002},
    journal = {Phys. Rept.},
    volume = {466},
    pages = {105--177},
    year = {2008}
}

@article{Fong:2012buy,
    author = {Fong, Chee Sheng and Nardi, Enrico and Riotto, Antonio},
    title = {{Leptogenesis in the Universe}},
    eprint = {1301.3062},
    archivePrefix = {arXiv},
    primaryClass = {hep-ph},
    doi = {10.1155/2012/158303},
    journal = {Adv. High Energy Phys.},
    volume = {2012},
    pages = {158303},
    year = {2012}
}

@article{Blanchet:2012bk,
    author = {Blanchet, Steve and Di Bari, Pasquale},
    title = {{The minimal scenario of leptogenesis}},
    eprint = {1211.0512},
    archivePrefix = {arXiv},
    primaryClass = {hep-ph},
    doi = {10.1088/1367-2630/14/12/125012},
    journal = {New J. Phys.},
    volume = {14},
    pages = {125012},
    year = {2012}
}

@article{Barbieri:1999ma,
    author = {Barbieri, Riccardo and Creminelli, Paolo and Strumia, Alessandro and Tetradis, Nikolaos},
    title = {{Baryogenesis through leptogenesis}},
    eprint = {hep-ph/9911315},
    archivePrefix = {arXiv},
    reportNumber = {SNS-PH-99-15, IFUP-TH-99-55},
    doi = {10.1016/s0550-3213(00)00011-0},
    journal = {Nucl. Phys. B},
    volume = {575},
    pages = {61--77},
    year = {2000}
}

@article{Abada:2006ea,
    author = {Abada, A. and Davidson, S. and Ibarra, A. and Josse-Michaux, F. -X. and Losada, M. and Riotto, A.},
    title = {{Flavour Matters in Leptogenesis}},
    eprint = {hep-ph/0605281},
    archivePrefix = {arXiv},
    reportNumber = {CERN-PH-TH-2006-093, DNI-UAN-06-97FT, IFT-UAM-CSIC-06-23, LPT-ORSAY-06-21, LYCEN-2006-07},
    doi = {10.1088/1126-6708/2006/09/010},
    journal = {JHEP},
    volume = {09},
    pages = {010},
    year = {2006}
}

@article{Nardi:2006fx,
    author = {Nardi, Enrico and Nir, Yosef and Roulet, Esteban and Racker, Juan},
    title = {{The Importance of flavor in leptogenesis}},
    eprint = {hep-ph/0601084},
    archivePrefix = {arXiv},
    doi = {10.1088/1126-6708/2006/01/164},
    journal = {JHEP},
    volume = {01},
    pages = {164},
    year = {2006}
}

@article{Giudice:2003jh,
    author = {Giudice, G. F. and Notari, A. and Raidal, M. and Riotto, A. and Strumia, A.},
    title = {{Towards a complete theory of thermal leptogenesis in the SM and MSSM}},
    eprint = {hep-ph/0310123},
    archivePrefix = {arXiv},
    reportNumber = {IFUP-TH-2003-37, CERN-TH-2003-240, IFUP-TH/2003-37 and CERN-TH/2003-240},
    doi = {10.1016/j.nuclphysb.2004.02.019},
    journal = {Nucl. Phys. B},
    volume = {685},
    pages = {89--149},
    year = {2004}
}

@article{Garbrecht:2018mrp,
    author = {Garbrecht, Bj{\"o}rn},
    title = {{Why is there more matter than antimatter? Calculational methods for leptogenesis and electroweak baryogenesis}},
    eprint = {1812.02651},
    archivePrefix = {arXiv},
    primaryClass = {hep-ph},
    reportNumber = {TUM-HEP-1177-18},
    doi = {10.1016/j.ppnp.2019.103727},
    journal = {Prog. Part. Nucl. Phys.},
    volume = {110},
    pages = {103727},
    year = {2020}
}

@article{Minkowski:1977sc,
    author = {Minkowski, Peter},
    title = {{$\mu \to e\gamma$ at a Rate of One Out of $10^{9}$ Muon Decays?}},
    reportNumber = {Print-77-0182 (BERN)},
    doi = {10.1016/0370-2693(77)90435-x},
    journal = {Phys. Lett. B},
    volume = {67},
    pages = {421--428},
    year = {1977}
}

@article{Gell-Mann:1979vob,
    author = {Gell-Mann, Murray and Ramond, Pierre and Slansky, Richard},
    title = {{Complex Spinors and Unified Theories}},
    eprint = {1306.4669},
    archivePrefix = {arXiv},
    primaryClass = {hep-th},
    reportNumber = {PRINT-80-0576},
    journal = {Conf. Proc. C},
    volume = {790927},
    pages = {315--321},
    year = {1979}
}

@article{Yanagida:1980xy,
    author = {Yanagida, Tsutomu},
    title = {{Horizontal Symmetry and Masses of Neutrinos}},
    reportNumber = {TU-80-208},
    doi = {10.1143/ptp.64.1103},
    journal = {Prog. Theor. Phys.},
    volume = {64},
    pages = {1103--1105},
    year = {1980}
}

@article{Mohapatra:1979ia,
    author = {Mohapatra, Rabindra N. and Senjanovic, Goran},
    title = {{Neutrino Mass and Spontaneous Parity Nonconservation}},
    reportNumber = {MDDP-TR-80-060, MDDP-PP-80-105, CCNY-HEP-79-10},
    doi = {10.1103/physrevlett.44.912},
    journal = {Phys. Rev. Lett.},
    volume = {44},
    pages = {912--915},
    year = {1980}
}

@article{Casas:2001sr,
    author = {Casas, J. A. and Ibarra, A.},
    title = {{Oscillating neutrinos and $\mu \to e, \gamma$}},
    eprint = {hep-ph/0103065},
    archivePrefix = {arXiv},
    reportNumber = {IEM-FT-211-01, OUTP-01-11P, IFT-UAM-CSIC-01-08},
    doi = {10.1016/s0550-3213(01)00475-8},
    journal = {Nucl. Phys. B},
    volume = {618},
    pages = {171--204},
    year = {2001}
}

@article{Kobayashi:1973fv,
    author = {Kobayashi, Makoto and Maskawa, Toshihide},
    title = {{CP Violation in the Renormalizable Theory of Weak Interaction}},
    reportNumber = {KUNS-242},
    doi = {10.1143/ptp.49.652},
    journal = {Prog. Theor. Phys.},
    volume = {49},
    pages = {652--657},
    year = {1973}
}

@article{Jarlskog:1985ht,
    author = {Jarlskog, C.},
    title = {{Commutator of the Quark Mass Matrices in the Standard Electroweak Model and a Measure of Maximal CP Nonconservation}},
    reportNumber = {USIP-85-14},
    doi = {10.1103/physrevlett.55.1039},
    journal = {Phys. Rev. Lett.},
    volume = {55},
    pages = {1039--1042},
    year = {1985}
}

@book{Branco:1999fs,
    author = {Branco, Gustavo C. and Lavoura, Luis and Silva, Joao P.},
    title = {{CP Violation}},
    doi = {10.1093/oso/9780198503996.001.0001},
    ISBN = {978-1-383-02075-5, 978-0-19-850399-6},
    publisher = {Oxford University Press},
    volume = {103},
    year = {1999}
}

@article{Dolgov:1982th,
    author = {Dolgov, A. D. and Linde, Andrei D.},
    title = {{Baryon Asymmetry in Inflationary Universe}},
    reportNumber = {ITEP-78-1982},
    doi = {10.1016/0370-2693(82)90292-1},
    journal = {Phys. Lett. B},
    volume = {116},
    pages = {329--334},
    year = {1982}
}

@article{Abbott:1982hn,
    author = {Abbott, L. F. and Farhi, Edward and Wise, Mark B.},
    title = {{Particle Production in the New Inflationary Cosmology}},
    reportNumber = {MIT-CTP-983},
    doi = {10.1016/0370-2693(82)90867-x},
    journal = {Phys. Lett. B},
    volume = {117},
    pages = {29--33},
    year = {1982}
}

@article{Traschen:1990sw,
    author = {Traschen, Jennie H. and Brandenberger, Robert H.},
    title = {{Particle Production During Out-of-equilibrium Phase Transitions}},
    reportNumber = {BROWN-HET-731},
    doi = {10.1103/physrevd.42.2491},
    journal = {Phys. Rev. D},
    volume = {42},
    pages = {2491--2504},
    year = {1990}
}

@article{Kofman:1994rk,
    author = {Kofman, Lev and Linde, Andrei D. and Starobinsky, Alexei A.},
    title = {{Reheating after inflation}},
    eprint = {hep-th/9405187},
    archivePrefix = {arXiv},
    reportNumber = {UH-IFA-94-35, SU-ITP-94-13, YITP-U-94-15},
    doi = {10.1103/physrevlett.73.3195},
    journal = {Phys. Rev. Lett.},
    volume = {73},
    pages = {3195--3198},
    year = {1994}
}

@article{Shtanov:1994ce,
    author = {Shtanov, Y. and Traschen, Jennie H. and Brandenberger, Robert H.},
    title = {{Universe reheating after inflation}},
    eprint = {hep-ph/9407247},
    archivePrefix = {arXiv},
    reportNumber = {BROWN-HET-957},
    doi = {10.1103/physrevd.51.5438},
    journal = {Phys. Rev. D},
    volume = {51},
    pages = {5438--5455},
    year = {1995}
}

@article{Kofman:1997yn,
    author = {Kofman, Lev and Linde, Andrei D. and Starobinsky, Alexei A.},
    title = {{Towards the theory of reheating after inflation}},
    eprint = {hep-ph/9704452},
    archivePrefix = {arXiv},
    reportNumber = {IFA-97-28, SU-ITP-97-18},
    doi = {10.1103/physrevd.56.3258},
    journal = {Phys. Rev. D},
    volume = {56},
    pages = {3258--3295},
    year = {1997}
}

@book{Kolb:1990vq,
    author = {Kolb, Edward W. and Turner, Michael S.},
    title = {{The Early Universe}},
    reportNumber = {FERMILAB-BOOK-1990-01},
    doi = {10.1201/9780429492860},
    ISBN = {978-0-429-49286-0, 978-0-201-62674-2},
    publisher = {Taylor and Francis},
    volume = {69},
    year = {2018},
    note = {Originally published in 1990}
}

@article{Giudice:2000ex,
    author = {Giudice, Gian Francesco and Kolb, Edward W. and Riotto, Antonio},
    title = {{Largest temperature of the radiation era and its cosmological implications}},
    eprint = {hep-ph/0005123},
    archivePrefix = {arXiv},
    reportNumber = {SNS-PH-00-05, FERMILAB-PUB-00-075-A, CERN-TH-2000-107},
    doi = {10.1103/physrevd.64.023508},
    journal = {Phys. Rev. D},
    volume = {64},
    pages = {023508},
    year = {2001}
}

@article{Allahverdi:2010xz,
    author = {Allahverdi, Rouzbeh and Brandenberger, Robert and Cyr-Racine, Francis-Yan and Mazumdar, Anupam},
    title = {{Reheating in Inflationary Cosmology: Theory and Applications}},
    eprint = {1001.2600},
    archivePrefix = {arXiv},
    primaryClass = {hep-th},
    doi = {10.1146/annurev.nucl.012809.104511},
    journal = {Ann. Rev. Nucl. Part. Sci.},
    volume = {60},
    pages = {27--51},
    year = {2010}
}

@article{Lozanov:2019jxc,
    author = {Lozanov, Kaloian D.},
    title = {{Lectures on Reheating after Inflation}},
    eprint = {1907.04402},
    archivePrefix = {arXiv},
    primaryClass = {astro-ph.CO},
    howpublished = {{arXiv:1907.04402}},
    month = {7},
    year = {2019}
}

@article{Planck:2018vyg,
    author = {Aghanim, N. and others},
    collaboration = {Planck},
    title = {{Planck 2018 results. VI. Cosmological parameters}},
    eprint = {1807.06209},
    archivePrefix = {arXiv},
    primaryClass = {astro-ph.CO},
    doi = {10.1051/0004-6361/201833910},
    journal = {Astron. Astrophys.},
    volume = {641},
    pages = {A6},
    year = {2020},
    note = {[Erratum: Astron.Astrophys. 652, C4 (2021)]}
}

@article{ParticleDataGroup:2024cfk,
    author = {Navas, S. and others},
    collaboration = {Particle Data Group},
    title = {{Review of particle physics}},
    doi = {10.1103/physrevd.110.030001},
    journal = {Phys. Rev. D},
    volume = {110},
    number = {3},
    pages = {030001},
    year = {2024}
}

@article{Garbrecht:2024ngz,
    author = {Garbrecht, Bj{\"o}rn and Wang, Edward},
    title = {{The neutrino mass bound from leptogenesis revisited}},
    eprint = {2411.09765},
    archivePrefix = {arXiv},
    primaryClass = {hep-ph},
    reportNumber = {TUM-HEP-1535/24},
    doi = {10.1007/jhep03(2025)008},
    journal = {JHEP},
    volume = {03},
    pages = {008},
    year = {2025}
}

@article{DOnofrio:2014rug,
    author = {D'Onofrio, Michela and Rummukainen, Kari and Tranberg, Anders},
    title = {{Sphaleron Rate in the Minimal Standard Model}},
    eprint = {1404.3565},
    archivePrefix = {arXiv},
    primaryClass = {hep-ph},
    doi = {10.1103/physrevlett.113.141602},
    journal = {Phys. Rev. Lett.},
    volume = {113},
    number = {14},
    pages = {141602},
    year = {2014}
}

@article{BhupalDev:2014pfm,
    author = {Bhupal Dev, P. S. and Millington, Peter and Pilaftsis, Apostolos and Teresi, Daniele},
    title = {{Flavour Covariant Transport Equations: an Application to Resonant Leptogenesis}},
    eprint = {1404.1003},
    archivePrefix = {arXiv},
    primaryClass = {hep-ph},
    reportNumber = {MAN-HEP-2014-01, IPPP-14-20, DCPT-14-40},
    doi = {10.1016/j.nuclphysb.2014.06.020},
    journal = {Nucl. Phys. B},
    volume = {886},
    pages = {569--664},
    year = {2014},
    note = {Corrigendum: Nucl. Phys. B 897 (2015) 749--756}
}

@article{Garny:2010nz,
    author = {Garny, M. and Hohenegger, A. and Kartavtsev, A.},
    title = {{Quantum corrections to leptogenesis from the gradient expansion}},
    eprint = {1005.5385},
    archivePrefix = {arXiv},
    primaryClass = {hep-ph},
    howpublished = {{arXiv:1005.5385}},
    month = {5},
    year = {2010}
}

@article{Esteban:2024eli,
    author = {Esteban, Ivan and Gonzalez-Garcia, M. C. and Maltoni, Michele and Martinez-Soler, Ivan and Pinheiro, Jo{\~a}o Paulo and Schwetz, Thomas},
    title = {{NuFit-6.0: updated global analysis of three-flavor neutrino oscillations}},
    eprint = {2410.05380},
    archivePrefix = {arXiv},
    primaryClass = {hep-ph},
    reportNumber = {IFT-UAM/CSIC-24-140, YITP-SB-2024-24, IPPP/24/64, IPPP/24/64, IFT-UAM/CSIC-24-140, YITP-SB-2024-24},
    doi = {10.1007/jhep12(2024)216},
    journal = {JHEP},
    volume = {12},
    pages = {216},
    year = {2024}
}

@article{Garbrecht:2011ctp,
    title = {{Effective theory of Resonant Leptogenesis in the Closed-Time-Path approach}},
    volume = {861},
    ISSN = {0550-3213},
    url = {http://dx.doi.org/10.1016/j.nuclphysb.2012.03.009},
    doi = {10.1016/j.nuclphysb.2012.03.009},
    number = {1},
    journal = {Nucl. Phys. B},
    publisher = {Elsevier BV},
    author = {Garbrecht, Bj{\"o}rn and Herranen, Matti},
    year = {2012},
    month = {Aug},
    pages = {17--52},
    eprint = {1112.5954},
    archivePrefix = {arXiv},
    primaryClass = {hep-ph}
}

@article{Beneke:2010ctp,
    title = {{Flavoured leptogenesis in the CTP formalism}},
    volume = {843},
    ISSN = {0550-3213},
    url = {http://dx.doi.org/10.1016/j.nuclphysb.2010.10.001},
    doi = {10.1016/j.nuclphysb.2010.10.001},
    number = {1},
    journal = {Nucl. Phys. B},
    publisher = {Elsevier BV},
    author = {Beneke, Martin and Garbrecht, Bj{\"o}rn and Fidler, Christian and Herranen, Matti and Schwaller, Pedro},
    year = {2011},
    month = {Feb},
    pages = {177--212},
    eprint = {1007.4783},
    archivePrefix = {arXiv},
    primaryClass = {hep-ph}
}

@article{Nardi:2007scatter,
    title = {{CP violation in scatterings, three body processes and the Boltzmann equations for leptogenesis}},
    volume = {09},
    ISSN = {1029-8479},
    url = {http://dx.doi.org/10.1088/1126-6708/2007/09/090},
    doi = {10.1088/1126-6708/2007/09/090},
    number = {09},
    journal = {JHEP},
    publisher = {Springer Science and Business Media LLC},
    author = {Nardi, Enrico and Racker, Juan and Roulet, Esteban},
    year = {2007},
    month = {sep},
    pages = {090},
    eprint = {0707.0378},
    archivePrefix = {arXiv},
    primaryClass = {hep-ph}
}

@article{Fong:2010gauge,
    title = {{CP violation from scatterings with gauge bosons in leptogenesis}},
    volume = {697},
    ISSN = {0370-2693},
    url = {http://dx.doi.org/10.1016/j.physletb.2011.02.025},
    doi = {10.1016/j.physletb.2011.02.025},
    number = {5},
    journal = {Phys. Lett. B},
    publisher = {Elsevier BV},
    author = {Fong, Chee Sheng and Gonzalez-Garcia, M.C. and Racker, J.},
    year = {2011},
    month = {Mar},
    pages = {463--470},
    eprint = {1010.2209},
    archivePrefix = {arXiv},
    primaryClass = {hep-ph}
}

@article{HahnWoernle:2009full,
    title = {{Full Boltzmann equations for leptogenesis including scattering}},
    volume = {08},
    ISSN = {1475-7516},
    url = {http://dx.doi.org/10.1088/1475-7516/2009/08/028},
    doi = {10.1088/1475-7516/2009/08/028},
    number = {08},
    journal = {JCAP},
    publisher = {IOP Publishing},
    author = {Hahn-Woernle, F. and Pl{\"u}macher, M. and Wong, Y. Y. Y.},
    year = {2009},
    month = {Aug},
    pages = {028},
    eprint = {0907.0205},
    archivePrefix = {arXiv},
    primaryClass = {hep-ph}
}

@article{Racker:2018nlo,
    author = {Racker, J.},
    title = {{Unitarity and CP violation in leptogenesis at NLO: general considerations and top Yukawa contributions}},
    eprint = {1811.00280},
    archivePrefix = {arXiv},
    primaryClass = {hep-ph},
    doi = {10.1007/jhep02(2019)042},
    journal = {JHEP},
    volume = {02},
    pages = {042},
    year = {2019}
}

@article{Falkowski:2011,
    author = {Falkowski, Adam and Ruderman, Joshua T. and Volansky, Tomer},
    title = {{Asymmetric Dark Matter from Leptogenesis}},
    eprint = {1101.4936},
    archivePrefix = {arXiv},
    primaryClass = {hep-ph},
    doi = {10.1007/jhep05(2011)106},
    journal = {JHEP},
    volume = {05},
    pages = {106},
    year = {2011}
}

@article{GreeneKofman:2000,
    author = {Greene, Patrick B. and Kofman, Lev},
    title = {{On the Theory of Fermionic Preheating}},
    eprint = {hep-ph/0003018},
    archivePrefix = {arXiv},
    primaryClass = {hep-ph},
    doi = {10.1103/physrevd.62.123516},
    journal = {Phys. Rev. D},
    volume = {62},
    pages = {123516},
    year = {2000}
}

@article{Branco:2001pq,
    author = {Branco, Gustavo C. and Morozumi, T. and Nobre, B. M. and Rebelo, M. N.},
    title = {{A Bridge between CP violation at Low Energies and Leptogenesis}},
    eprint = {hep-ph/0107164},
    archivePrefix = {arXiv},
    doi = {10.1016/s0550-3213(01)00425-4},
    journal = {Nucl. Phys. B},
    volume = {617},
    pages = {475--492},
    year = {2001}
}

@article{Branco:2005jr,
    author = {Branco, Gustavo C. and Rebelo, M. N. and Silva-Marcos, J. I.},
    title = {{Leptogenesis, Yukawa textures and weak-basis invariants}},
    eprint = {hep-ph/0510412},
    archivePrefix = {arXiv},
    doi = {10.1016/j.physletb.2005.11.067},
    journal = {Phys. Lett. B},
    volume = {633},
    pages = {345--354},
    year = {2006}
}

@article{Choubey:2010up,
    author = {Choubey, Sandhya and King, Stephen F. and Mitra, Manimala},
    title = {{Vanishing of the CP asymmetry in leptogenesis due to form dominance}},
    eprint = {1004.3756},
    archivePrefix = {arXiv},
    primaryClass = {hep-ph},
    doi = {10.1103/physrevd.82.033002},
    journal = {Phys. Rev. D},
    volume = {82},
    pages = {033002},
    year = {2010}
}

@article{Chen:2016ptr,
    author = {Chen, Peng and Ding, Gui-Jun and King, Stephen F.},
    title = {{Leptogenesis and residual CP symmetry}},
    eprint = {1602.03873},
    archivePrefix = {arXiv},
    primaryClass = {hep-ph},
    doi = {10.1007/jhep03(2016)206},
    journal = {JHEP},
    volume = {03},
    pages = {206},
    year = {2016}
}

@article{Li:2017zmk,
    author = {Li, Cai-Chang and Ding, Gui-Jun},
    title = {{Implications of residual CP symmetry for leptogenesis in a model with two right-handed neutrinos}},
    eprint = {1701.08508},
    archivePrefix = {arXiv},
    primaryClass = {hep-ph},
    doi = {10.1103/physrevd.96.075005},
    journal = {Phys. Rev. D},
    volume = {96},
    number = {7},
    pages = {075005},
    year = {2017}
}

@article{HahnWoernle:2008reheating,
    author = {Hahn-Woernle, F. and Pl{\"u}macher, M.},
    title = {{Effects of reheating on leptogenesis}},
    eprint = {0801.3972},
    archivePrefix = {arXiv},
    primaryClass = {hep-ph},
    doi = {10.1016/j.nuclphysb.2008.07.032},
    journal = {Nucl. Phys. B},
    volume = {806},
    pages = {68--83},
    year = {2009}
}

@article{Datta:2023flavor,
    author = {Datta, Arghyajit and Roshan, Rishav and Sil, Arunansu},
    title = {{Flavor leptogenesis during the reheating era}},
    eprint = {2301.10791},
    archivePrefix = {arXiv},
    primaryClass = {hep-ph},
    doi = {10.1103/physrevd.108.035029},
    journal = {Phys. Rev. D},
    volume = {108},
    number = {3},
    pages = {035029},
    year = {2023}
}

@article{Zhang:2023nonthermal,
    author = {Zhang, Xinyi},
    title = {{Towards a systematic study of non-thermal leptogenesis from inflaton decays}},
    eprint = {2311.05824},
    archivePrefix = {arXiv},
    primaryClass = {hep-ph},
    doi = {10.1007/jhep05(2024)147},
    journal = {JHEP},
    volume = {05},
    pages = {147},
    year = {2024}
}

@article{Falkowski:2017lit,
    author = "Falkowski, Adam and Kuflik, Eric and Levi, Noam and Volansky, Tomer",
    title = "{Light Dark Matter from Leptogenesis}",
    eprint = "1712.07652",
    archivePrefix = "arXiv",
    primaryClass = "hep-ph",
    doi = "10.1103/PhysRevD.99.015022",
    journal = "Phys. Rev. D",
    volume = "99",
    number = "1",
    pages = "015022",
    year = "2019"
}

@article{HerreroGarcia:2024seesaw,
    author = "Herrero-Garcia, Juan and Landini, Giacomo and Yanagida, Tsutomu T.",
    title = "{Dark matter in the high-scale seesaw leptogenesis paradigm}",
    eprint = "2411.03452",
    archivePrefix = "arXiv",
    primaryClass = "hep-ph",
    doi = "10.1103/PhysRevD.111.075033",
    journal = "Phys. Rev. D",
    volume = "111",
    number = "7",
    pages = "075033",
    year = "2025"
}

@article{Escudero:2019decay,
    author = "Escudero, Miguel and Fairbairn, Malcolm",
    title = "{Cosmological Constraints on Invisible Neutrino Decays Revisited}",
    eprint = "1907.05425",
    archivePrefix = "arXiv",
    primaryClass = "hep-ph",
    reportNumber = "KCL-2019-57",
    doi = "10.1103/PhysRevD.100.103531",
    journal = "Phys. Rev. D",
    volume = "100",
    number = "10",
    pages = "103531",
    year = "2019"
}

@article{Barenboim:2020decay,
    author = "Barenboim, Gabriela and Chen, Joe Zhiyu and Hannestad, Steen and Oldengott, Isabel M. and Tram, Thomas and Wong, Yvonne Y. Y.",
    title = "{Invisible neutrino decay in precision cosmology}",
    eprint = "2011.01502",
    archivePrefix = "arXiv",
    primaryClass = "astro-ph.CO",
    doi = "10.1088/1475-7516/2021/03/087",
    journal = "JCAP",
    volume = "03",
    pages = "087",
    year = "2021"
}

@article{Chen:2022lifetime,
    author = "Chen, Joe Zhiyu and Oldengott, Isabel M. and Pierobon, Giovanni and Wong, Yvonne Y. Y.",
    title = "{Weaker yet again: mass spectrum-consistent cosmological constraints on the neutrino lifetime}",
    eprint = "2203.09075",
    archivePrefix = "arXiv",
    primaryClass = "hep-ph",
    doi = "10.1140/epjc/s10052-022-10518-3",
    journal = "Eur. Phys. J. C",
    volume = "82",
    number = "7",
    pages = "640",
    year = "2022"
}

@article{Kaiser:1993lasing,
    author = "Kaiser, Nick and Malaney, R. A. and Starkman, G. D.",
    title = "{Neutrino lasing in the early universe}",
    eprint = "hep-ph/9302261",
    archivePrefix = "arXiv",
    reportNumber = "CITA-93-7",
    doi = "10.1103/PhysRevLett.71.1128",
    journal = "Phys. Rev. Lett.",
    volume = "71",
    pages = "1128--1131",
    year = "1993"
}

@article{Starkman:1993lasing,
    author = "Starkman, G. D. and Kaiser, Nick and Malaney, R. A.",
    title = "{Mixed dark matter from neutrino lasing}",
    eprint = "astro-ph/9312020",
    archivePrefix = "arXiv",
    reportNumber = "CITA-93-49",
    doi = "10.1086/174700",
    journal = "Astrophys. J.",
    volume = "434",
    pages = "12--23",
    year = "1994"
}

@article{PerezGonzalez:2026thermal,
    author = "Perez-Gonzalez, Yuber F. and Sen, Manibrata and Tangarife, Walter",
    title = "{Bose-enhanced Neutrino Decays in a Thermal Medium}",
    eprint = "2606.14441",
    archivePrefix = "arXiv",
    primaryClass = "hep-ph",
    month = "6",
    year = "2026"
}

@article{Bento:2001hidden,
    author = "Bento, Luis and Berezhiani, Zurab",
    title = "{Leptogenesis via Collisions: Leaking Lepton Number to the Hidden Sector}",
    eprint = "hep-ph/0107281",
    archivePrefix = "arXiv",
    reportNumber = "CFNUL-01-04, DFAQ-2001-05-TH",
    doi = "10.1103/PhysRevLett.87.231304",
    journal = "Phys. Rev. Lett.",
    volume = "87",
    pages = "231304",
    year = "2001"
}

@article{Blazek:2025effective,
    author = "Bla{\v z}ek, Tom{\'a}{\v s} and Mat{\'a}k, Peter and Ramaj, J{\'a}n and Sabov{\'a}, Martina",
    title = "{Minimal effective theory for leptogenesis, dark matter, and neutrino masses}",
    eprint = "2504.15164",
    archivePrefix = "arXiv",
    primaryClass = "hep-ph",
    doi = "10.1140/epjc/s10052-025-14543-w",
    journal = "Eur. Phys. J. C",
    volume = "85",
    number = "7",
    pages = "801",
    year = "2025"
}

@article{Kiessig:2010thermal,
    author = {Kie{\ss}ig, Clemens P. and Pl{\"u}macher, Michael and Thoma, Markus H.},
    title = {{Decay of a Yukawa fermion at finite temperature and applications to leptogenesis}},
    eprint = {1003.3016},
    archivePrefix = {arXiv},
    primaryClass = {hep-ph},
    doi = {10.1103/PhysRevD.82.036007},
    journal = {Phys. Rev. D},
    volume = {82},
    pages = {036007},
    year = {2010}
}

@article{Du:2021distribution,
    author = {Du, Yong and Huang, Fei and Li, Hao-Lin and Li, Yuan-Zhen and Yu, Jiang-Hao},
    title = {{Revisiting dark matter freeze-in and freeze-out through phase-space distribution}},
    eprint = {2111.01267},
    archivePrefix = {arXiv},
    primaryClass = {hep-ph},
    doi = {10.1088/1475-7516/2022/04/012},
    journal = {JCAP},
    volume = {04},
    pages = {012},
    year = {2022}
}

@article{Mukaida:2015density,
    author = {Mukaida, Kyohei and Yamada, Masaki},
    title = {{Thermalization process after inflation and effective potential of scalar field}},
    eprint = {1506.07661},
    archivePrefix = {arXiv},
    primaryClass = {hep-ph},
    doi = {10.1088/1475-7516/2016/02/003},
    journal = {JCAP},
    volume = {02},
    pages = {003},
    year = {2016}
}

@article{Barreiros:2022fpi,
    author = {Barreiros, D. M. and C{\^a}mara, H. B. and Felipe, R. G. and Joaquim, F. R.},
    title = {{Scalar-singlet assisted leptogenesis with CP violation from the vacuum}},
    eprint = {2211.00042},
    archivePrefix = {arXiv},
    primaryClass = {hep-ph},
    doi = {10.1007/JHEP01(2023)010},
    journal = {JHEP},
    volume = {01},
    pages = {010},
    year = {2023}
}

@article{Coleman:1973jx,
    author = {Coleman, Sidney R. and Weinberg, Erick J.},
    title = {{Radiative Corrections as the Origin of Spontaneous Symmetry Breaking}},
    doi = {10.1103/PhysRevD.7.1888},
    journal = {Phys. Rev. D},
    volume = {7},
    pages = {1888--1910},
    year = {1973}
}

@inproceedings{Ramond:1979family,
    author = {Ramond, Pierre},
    title = {{The Family Group in Grand Unified Theories}},
    booktitle = {{International Symposium on Fundamentals of Quantum Theory and Quantum Field Theory}},
    eprint = {hep-ph/9809459},
    archivePrefix = {arXiv},
    reportNumber = {CALT-68-709},
    note = {Invited talk at the Sanibel Symposia, February 1979; retrospective preprint posted in 1998},
    year = {1979}
}

@article{Enqvist:2014singlet,
    author = {Enqvist, Kari and Nurmi, Sami and Tenkanen, Tommi and Tuominen, Kimmo},
    title = {{Standard Model with a real singlet scalar and inflation}},
    eprint = {1407.0659},
    archivePrefix = {arXiv},
    primaryClass = {astro-ph.CO},
    doi = {10.1088/1475-7516/2014/08/035},
    journal = {JCAP},
    volume = {08},
    pages = {035},
    year = {2014}
}

@article{Enqvist:2012spectator,
    author = {Enqvist, Kari and Lerner, Rose N. and Taanila, Olli and Tranberg, Anders},
    title = {{Spectator field dynamics in de Sitter and curvaton initial conditions}},
    eprint = {1205.5446},
    archivePrefix = {arXiv},
    primaryClass = {astro-ph.CO},
    doi = {10.1088/1475-7516/2012/10/052},
    journal = {JCAP},
    volume = {10},
    pages = {052},
    year = {2012}
}

@article{Antusch:2009nonunitary,
    author = {Antusch, Stefan and Blanchet, Steve and Blennow, Mattias and Fernandez-Martinez, Enrique},
    title = {{Non-unitary Leptonic Mixing and Leptogenesis}},
    eprint = {0910.5957},
    archivePrefix = {arXiv},
    primaryClass = {hep-ph},
    doi = {10.1007/JHEP01(2010)017},
    journal = {JHEP},
    volume = {01},
    pages = {017},
    year = {2010}
}

@article{LopezNacir:2014hartree,
    author = {L{\'o}pez Nacir, Diana L. and Mazzitelli, Francisco D. and Trombetta, Leonardo G.},
    title = {{Hartree approximation in curved spacetimes revisited II: The semiclassical Einstein equations and de Sitter self-consistent solutions}},
    eprint = {1401.6094},
    archivePrefix = {arXiv},
    primaryClass = {hep-th},
    doi = {10.1103/PhysRevD.89.084013},
    journal = {Phys. Rev. D},
    volume = {89},
    pages = {084013},
    year = {2014}
}
\end{document}